\documentstyle[eqsecnum,aps,epsf]{revtex}
\begin{document}   
    
\newcommand{\dsTx}{\mbox{\,$\Delta\sigma_{\rm \small T}$}}
\newcommand{\dsT}{\mbox{\,$\Delta\sigma_{\rm \small T}$\ }}
\newcommand{\dsLx}{\mbox{\,$\Delta\sigma_{\rm \small L}$}}
\newcommand{\dsL}{\mbox{\,$\Delta\sigma_{\rm \small L}$\ }}
\newcommand{\sigtotx}{\mbox{\,$\sigma_{\rm tot}$}}
\newcommand{\sigtot}{\mbox{\,$\sigma_{\rm tot}$\ }}
\newcommand{\grad}{\mbox{$^{\circ}$}\ }
\newcommand{\ppp}{$pp \rightarrow pp\pi^0\,$}
\newcommand{\ppn}{$pp \rightarrow pn\pi^+\,$}
\newcommand{\ccq}{\mbox{$(3\cos^2\theta_{\pi} - 1)$}}
\newcommand{\ccp}{(3\cos^2\theta_p - 1)}
\newcommand{\sinq}{\sin\theta_{\pi}}
\newcommand{\sinp}{\sin\theta_p}
\newcommand{\cosq}{\cos\theta_{\pi}}
\newcommand{\cosp}{\cos\theta_p}
\newcommand{\stq}{\sin 2\theta_{\pi}}
\newcommand{\stp}{\sin 2\theta_p}
\newcommand{\ssq}{\sin^2\theta_{\pi}}
\newcommand{\ssp}{\sin^2\theta_p}   
\draft\tighten                                 
\preprint{}
\title{Preprint 8/10/2001 \\
\    \\
\Large\bf Spin correlations in 
${\boldmath \rm \vec{p}\vec{p}\rightarrow pn\pi^{+}}$
	pion production near threshold}
\author {{W.W. Daehnick},\footnote{e-mail :
	 daehnick@pitt.edu.} 
	 Swapan K. Saha,\footnote{permanent address: Bose Institute,
         Calcutta 700009, India.} and R.W. Flammang,\footnote{present 
         address: Westinghouse Nuclear, Pittsburgh, PA.}}
\address{Department of Physics and Astronomy, University of Pittsburgh,
         Pittsburgh, PA 15260}
\author {H.O. Meyer, J.T. Balewski, R.E. Pollock, B. von Przewoski,
         T. Rinckel,\\
	 P. Th\"orngren-Engblom,\footnote{present 
         address: Dept. of Radiation Science, Uppsala, Sweden.}
	 and A. Wellinghausen ,\footnote{present address:
	 e-mail: arne.wellinghausen@gmx.net.}}
\address{Department of Physics and Cyclotron Facility, Indiana University,
         Bloomington, IN 47405}
\author {B. Lorentz \footnote[6]{present address: Institut f\"ur 
         Kernphysik, Forschungszentrum J\"ulich, 52425 J\"ulich, Germany.},
	 F. Rathmann,\footnotemark[6]{}
	         B. Schwartz, and T. Wise}
\address{University of Wisconsin-Madison, Madison, WI, 53706}
\author {P.V. Pancella}
\address{Western Michigan University, Kalamazoo, MI, 49008}

\maketitle
\vspace{0.5cm}
\vspace{0.5cm}

\begin{abstract}

A first measurement of longitudinal as well as transverse spin 
 correlation coefficients for the reaction 
 {$\vec{p}\vec{p}\rightarrow pn\pi^+$} was made using  
a polarized proton target and a polarized proton beam.  
We report kinematically complete measurements for this reaction 
at 325, 350, 375 and 400 MeV  beam energy. The spin correlation 
coefficients $A_{xx}+A_{yy}, A_{xx}-A_{yy}, A_{zz}, A_{xz},$ 
and the analyzing power $A_{y},$ as well as angular distributions 
for $\sigma(\theta_{\pi})$ and the polarization 
 observables $A_{ij}(\theta_{\pi})$ were extracted. 
Partial wave cross sections for dominant transition channels were 
obtained from a partial wave analysis that included the transitions 
with final state angular momenta of $l\leq 1$.  
The  measurements of the ${\vec{p}\vec{p}\rightarrow pn\pi^{+}}$
 polarization observables are compared with the predictions from the 
 J\"ulich meson exchange model. The agreement is very good
 at 325 MeV, but it deteriorates increasingly for the higher 
 energies. At all energies agreement with the model is better than 
for the reaction  ${\vec{p}\vec{p}\rightarrow pp\pi^{0}}$.
\end{abstract}

\input{epsf}

\twocolumn
\section{Introduction}

The pion-nucleon interaction has provided 
increasingly sensitive tests of nuclear theory. One of the 
challenges yet to be met is to understand the 
 polarization observables for pion production 
in $\vec{p} \vec{p}$ collisions. This is
especially interesting near threshold where 
few partial waves contribute and where calculations should 
be more manageable and  more conclusive.

After the initial theoretical work in the fifties 
by Gell-Mann and Watson \cite{gell-mann} 
and Rosenfeld \cite{rosenfeld}, more than a decade elapsed
before explicit {${p}{p}\rightarrow pp\pi^0$}
and {${p}{p}\rightarrow pn\pi^+$} cross sections for Ss 
($l_{NN}=0,\ l_{\pi}=0$) transitions were predicted by Koltun 
and Reitan \cite{koltun} in 1966
and by Schillaci, Silbar and Young \cite{schillaci} in 1969.
When the small cross sections very close to threshold could finally be 
measured twenty years later \cite{meyer90,daehnick95}, it 
turned out that these calculations had missed the true cross 
sections by factors up to five. This realization 
has spurred much new theoretical research.

To date the J\"ulich meson exchange model 
\cite{haidenbauer96,hanhart97,hanhart98a,hanhart98b,hanhart2000} 
has yielded the most successful
calculations. This model represents a much advanced development 
of the approach of Ref.\cite{koltun} and  
builds on the insights of the 1990's (e.g., by Lee and Riska 
\cite{riska} and many others). It  has 
permitted detailed calculations beyond $l_{\pi}=0$ transitions and 
provides analyzing powers and spin correlation coefficients
for the near-threshold region. 
The J\"ulich model incorporates all the basic diagrams,
realistic final state interactions,
off-shell effects, contributions from the delta resonance and
the exchange of heavier mesons.
With the exception of the heavy meson exchange term 
there are no adjustable parameters. At this time it is the only 
model with predictions that can be compared to our measurements.
However, the J\"ulich model
does not account for quark degrees of 
freedom, the potential study of which  had motivated 
our experiment initially. 

Ideally, one would interpret the basic pion production 
reactions in a framework compatible with QCD, e.g., 
calculations using chiral perturbation theory ($\chi PT$). 
However, with one exception \cite{hanhart2000b}, 
the $\chi PT$ calculations published to date are 
still restricted to $l_{\pi}=0$. Moreover, for all three 
$pp\rightarrow X\pi$ reactions, the $\chi PT$ cross sections
remain a factor of two or more below experiment \cite{darocha}. 
This shortcoming may be attributable to the difficulties of 
$\chi PT$ for momentum transfers larger than $m_{\pi}$. The
 $\chi PT$ calculations published to date
are best viewed as work in progress
\cite{blankleider}.

Calculations and experiments very close to threshold 
require great care. For {${p}{p}\rightarrow pn\pi^+$} Ss
transitions ($l_{\pi}=0$, $l_{pn}=0$) only one amplitude is calculated 
and the angular dependence is trivial. 
However, the near-threshold cross section and its 
energy dependence are significantly modified 
by ``secondary'' effects, such as final state interactions 
that are particularly important for $l_{pn}=0$.

Measurements very close to threshold can present difficulties 
because the cross sections are small, of the order of 1 $\mu b$, 
and change rapidly with energy. The energy of the interacting 
nucleons for reactions very close to threshold
must be precisely known and maintained.
At IUCF this has been accomplished by the use of
a very thin internal target 
and the precise beam energy control of the Cooler (storage) Ring.  
The IUCF Cooler also generates low background.

The earliest studies of {${p}{p}\rightarrow pn\pi^+$} very 
close to threshold \cite{daehnick95,hardie,daehnick98,flammang}
had available a stored IUCF beam of  
$\le 50 \mu A$. They used an unpolarized gas jet target and
measured cross sections and analyzing powers from 293 MeV  
(i.e., 0.7 MeV above the $\pi^{+}$ production threshold) 
to 330 MeV. These experiments deduced cross sections for 
Ss pion production
very close to threshold. As long as Ss production of 
pions strongly dominated, analyzing powers also provided 
information for Sp ($l_{\pi}=1$) admixtures \cite{flammang}.
At 325 MeV and above, higher partial waves enter significantly, 
but the larger cross sections make it 
practical to explore analyzing powers and spin 
correlation coefficients, which 
allow a  much more detailed comparison of theory and 
experiment.
At the upgraded IUCF Cooler Ring, an intense
polarized proton beam with a large longitudinal component
and an efficient windowless polarized 
hydrogen target now permit measurements of 
all spin correlations coefficients for 
$\vec{p}\vec{p}\rightarrow pn\pi^+$. 
Some initial results for transverse spin correlations 
have been reported in \cite{saha99}.

The goal of the present study is to quantify the 
growing importance of higher 
partial waves (Sp, Ps, Pp, and, potentially, Sd transitions)
by measuring analyzing powers and spin correlation 
coefficients as a function of energy. These polarization
observables are sensitive indicators of the 
reaction mechanism and the contributing partial waves
\cite{flammang,meyer00,knutsen} and are a powerful tool in 
determining transition amplitudes empirically.
In this experiment we measured
$A_{xx}+A_{yy},\ A_{xx}-A_{yy},\ A_{zz},\ A_{xz}$, as well as 
the polarization observables $A_{y}$ and $A_{z}$ for 
the energy region 325 to 400 MeV. 

\section{Experiment}
\subsection{Experimental considerations}

The Cooler Ring of the Indiana University
Cyclotron Facility (IUCF) produces 
protons of energies up to 500 MeV with
polarization of $\rm P \approx 0.65$, low 
emittance and low background. This permits 
in-beam experiments of reactions with microbarn 
cross sections. 
The improvement of beam intensity at IUCF over time now 
 allows the  use of very thin polarized targets.
During the {$\vec{p}\vec{p}\rightarrow pn\pi^+$} 
experiment typical intensities of the 
stored polarized beam ranged from $100 - 300 \mu A$.

The apparatus for polarized internal target experiments (PINTEX)  
makes use of a windowless target cell continuously filled by a 
polarized atomic Hydrogen beam.
The measurements cycle through a full set of relative 
beam and target spin alignments. The technical aspects of beam 
preparation, electronics, and target and detector 
properties have been reported previously in \cite{rinckel}. 
As is customary, the beam is defined to travel in 
the positive z direction, y is vertical and x completes a right 
handed coordinate system.
Below is a brief review of parameters pertinent to the data analysis.
%
\vspace{-3cm}
\hspace{-0.5cm}
\begin{figure}[htb]
\epsfysize=9.cm
\centerline{
\epsfbox{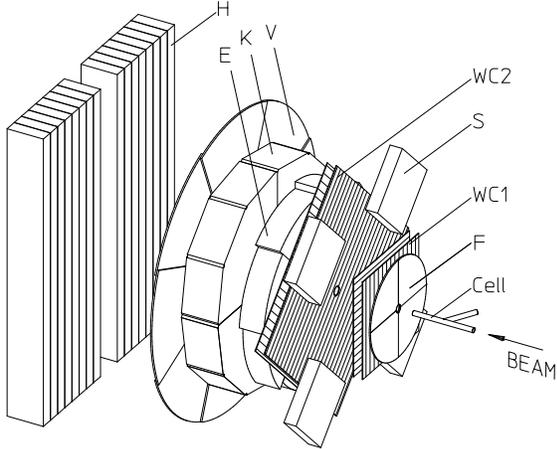}
}
\vspace{-0.0cm}
\caption{The PINTEX detector for the experiment: 
F is a thin timing detector. S labels one of the four detectors for 
the elastic pp scattering monitor. WC1 and WC2 are wire
chambers. E and K are segmented plastic scintillator stacks that 
determine the energy of the  charged reaction products. 
V is the charged particle veto detector, and H is the neutron 
hodoscope.} 
\label{fig:Fig1}
\end{figure}
In the experiment we used the Madison atomic beam target 
with a  storage cell of very low mass\cite{wise,ross}. 
The storage cell had a length of 25 cm  
and a diameter of 1.2 cm. This open-ended cylindrical 
cell produces a triangular shape of the target density distribution
with its maximum at the center (z=0). It was made of a thin  
(25 $\mu m$) aluminum foil to keep to a minimum background 
events caused by the beam halo. Teflon coating was used to inhibit 
depolarization of the target atoms.
Sets of orthogonal holding coils surround the storage cell.
The coils are used to align the polarized hydrogen 
atoms in the $\pm x, \pm y$, and $\pm z$  directions. 
Typical target polarizations were Q = 0.75 and the approximate 
target density was  $\rm 1.4 \times 10^{13}\ atoms/cm^{2}$. 

The target spin alignment can be changed in less than 10 ms.
During runs the target polarization direction was changed every 2 sec
and followed the sequence $\pm x,\pm y$, and $\pm z $.
Each data taking cycle had constant beam polarization 
and was set to last  5-8 minutes, after
which the remaining beam was discarded. 
The beam polarization was reversed with each new 
cycle to minimize the effect of apparatus asymmetries. In the first 
phase of the experiment (run a) the beam spin directions 
were alternated between +y and -y.
In the more recent runs (b) solenoid spin rotators were used to give 
the beam spin a large longitudinal component. This spin rotation 
was energy dependent and produced 
roughly equal longitudinal (z) and vertical (y) spin components and a 
very small component in the (x) direction, as shown in Table I.

Elastic $\vec{p} \vec{p}$ scattering was used to measure and 
monitor the three beam polarization components as well as the
luminosity. Elastic protons were detected with four plastic
scintillators mounted at $\theta= 45^{\circ}$, with  
$\phi=\pm 45^{\circ}$ and  $\pm 135^{\circ}$.
Coincident protons striking these monitor detectors 
(labeled S in Fig. 1) pass through wire 
chamber 1, so the needed tracking information is available.
The product PQ of beam polarization (P) and target polarization (Q)
was deduced from the large known  spin 
correlation $A_{xx} - A_{yy}$ in elastic scattering \cite{przewoski98}.
A 3D sketch of the detector system is shown in Fig.1.

The reaction pions in this study had lab energies from 0.1 
to 120.5 MeV and were emitted at polar lab
angles from $0^{\circ}$ to $180^{\circ}$. 
By contrast, the reaction nucleons remain constrained by 
kinematics to forward angles below $31.2 ^{\circ}$ and to 
lab energies from 20.8 to 227.9 MeV. This range 
of angles and energies affects 
the choice of detectors that can be employed. If both outgoing 
nucleons are protons as in {$\vec{p}\vec{p}\rightarrow pp\pi^0$}
one can ignore the pion and 
use a moderate size forward detector to intercept almost all 
ejectiles of interest \cite{rinckel}. This procedure was used for the 
simultaneously measured {$\vec{p}\vec{p}\rightarrow pp\pi^0$} 
reaction \cite{meyer01}. The corresponding procedure for 
{$\vec{p}\vec{p}\rightarrow pn\pi^+$} is to mount a large area 
neutron hodoscope behind the proton detectors and determine 
the  energies of the detected neutrons by time of flight. 
The construction and operation of the neutron hodoscope has 
been described previously in \cite{daehnick92}.
%
\vspace{-0.5cm}
\hspace{-1.cm}
\begin{figure}[htb]
\epsfysize=7.cm
\centerline{
\epsfbox{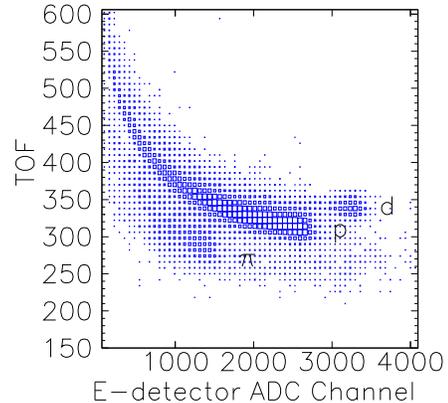}
}
\vspace{-.75cm}
\caption{Raw ejectile time of flight (ADC channels) vs. energy 
deposited in the E detector by pions, protons and deuterons. 
Triggers from two charged particles 
tracks and (no K) were a prerequisite. 
This spectrum was used for identification of protons and pions.}
\label{fig:et}
\end{figure}
All detectors are segmented because the energies of the coincident 
reaction particles need to be measured independently.
Monte Carlo calculations suggest that eight 
$\Delta \phi$ segments are sufficient because 
of the tendency of the ejectiles to have significantly 
different azimuthal angles. The  K 
detector was needed to obtain the necessary stopping power 
for the more energetic pions and protons.
Identification of the charged particles was usually 
accomplished by their time of flight vs. energy correlation, where 
the start signal was supplied by the F detector and the stop
signal was provided by the E detector. The pion and 
proton distributions were generally well separated.
Fig.~2 shows a typical particle ID spectrum for accepted 
$p\pi^{+}$ coincidence events.

\begin{figure}[htb]
\epsfysize=6.5cm
\centerline{
\epsfbox{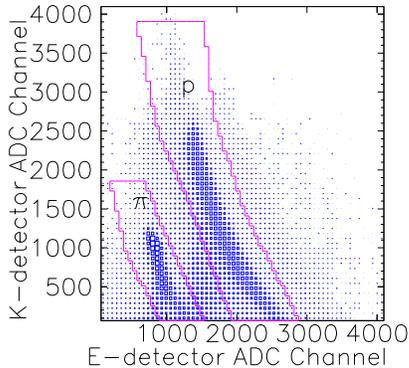}
}
\vspace{-0.5cm}
\caption{Particle identification cuts at 375 MeV for energetic ejectiles 
based on energy deposited in the K detector as function of energy loss 
in the E detector. Acceptable events had to be inside the regions 
outlined.}
\label{fig:ek}
\end{figure}

The more energetic ejectiles stop in the K detector, and superior
particle identification is obtained by comparing the energies deposited
in the K vs. the E detectors, as seen in Fig.~\ref{fig:ek}.
\vspace{-0.8cm}
\hspace{-1.cm}
We measured the polarization observables $A_{ij}$ in two different ways: 

1) by measuring the pions directly, in coincidence with protons (p+$\pi$ 
method), and

2) by reconstructing pion momenta from
the measured proton and neutron momenta (p+n method). 

The first method had the advantage of simplicity and high count rate,
but we cannot measure pions at large angles due to the
limited detector size. Therefore the spin dependent
cross section ratios could be compared with theoretical 
spin correlation coefficients only at forward angles.
The second method is free
from this limitation, but at the cost of the low neutron 
detection efficiency and therefore much lower statistics.

\subsection {Measurement of p + $\pi^{+}$ coincidences}

We accept events with two charged reaction particles (p and $\pi^{+}$) 
in coincidence. They must show separate tracks in the wire chambers 
WC1, WC2, trigger separate sections of the E detectors, 
and at least one section of the F 
detector, but not the veto scintillator (V).
The trajectories of the protons and pions are deduced from the wire 
chamber position readings. Their angular resolution was limited 
primarily by multiple scattering in the 1.5 mm thick F detector and in the 
0.18 mm thick stainless steel exit foil. Approximate angular resolutions 
(in the lab system) are 
$\sigma = 0.5^{\circ}$ for protons and $1.0^{\circ}$ for pions.
This resolution was fully sufficient for the angular 
variations expected. 

The good intrinsic angular resolution of the wire 
chambers was used to check the consistency of the pion and neutron 
position readouts by tracing elastic protons to the hodoscope bars and 
comparing predicted and observed position readings. 
It was found that from run to run the beam 
axis and the detector symmetry axis could differ slightly in direction 
and also in their relative x and y coordinates at z=0 (the target center).
We could also cross check the nominal z separation of the wire chambers
since the separation and location of the 
hodoscope bars was fixed and well known.
Small corrections of 1-3 mm had to be applied in software to the detector 
positions.  After such corrections the remaining systematic 
angular error of the measured polar angles is about $0.04^{\circ}$.

Charged particles that do not stop in or before the K detector 
trigger the V detector and are tagged as likely elastic events 
and generally vetoed. At 400 MeV we reach the design 
limit of the charged particle detectors and the veto detector begins 
to see (and reject) the most energetic pions at small angles.
In deducing the energy spectra account was taken of the differing 
non-linearity of light production for protons and pions
by the plastic scintillators,
as well as of energy losses in the exit foil, the F
detector, air and other materials between the scintillators. 
After calibrations of all detector segments the detector stack
provided an energy resolution for typical reaction protons
and pions of about $\Delta E/E = .09$ (FWHM).

\begin{figure}[htb]
\epsfysize=6.5cm
\centerline{
\epsfbox{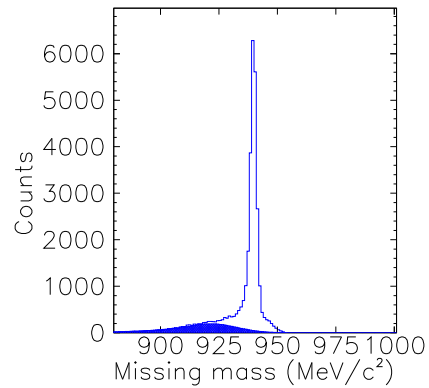}
}
\vspace{-0.5cm}
\caption{Neutron missing mass reconstructed from measured pion and 
proton momenta. The background spectrum shown (dark area)
results if the atomic hydrogen in 
the target cell is replaced by Nitrogen gas.}
\label{fig:nmm}
\end{figure}

The missing mass spectra contain a background continuum 
(see Fig.~\ref{fig:nmm}), which at higher beam energies 
stretches slightly beyond the missing mass peak.
%
\vspace{-0.5cm}
\hspace{-1.cm}
The trajectory traceback indicates that this background is primarily 
caused by beam halo hitting the Al and teflon components of 
the target cell. Without target gas (and the beam heating
normally produced by it) almost no background 
is seen. 
In order to  obtain a realistic background shape
near and below the missing mass peak the target cell was
filled with $N_{2}$. This gas will produce some background 
of its own, but just as importantly it heats the circulating beam 
(as the Hydrogen gas would) and reproduces the ordinary beam 
halo. We found that the  ``$N_{2}$ spectra''
seen after the common software cuts looked identical 
to the background ``tail'' in the Hydrogen missing mass spectra.
Therefore, $N_{2}$ spectra were measured with good statistics, 
and their shape was later used to correct for background 
under the missing mass peak.
Our statistically most accurate measurements were obtained in the 
p+$\pi$ mode, i.e., by observing pions and protons in coincidence.  

\subsection {Measurement of p+n coincidences}

Reaction neutrons in coincidence with protons were detected in a 
large hodoscope consisting of 16 long plastic scintillator bars.   
The bars were placed symmetrically about the beam direction in a 
plane defined by z=1.48 m. They were 15 cm deep and 
mounted so that their dimension in the y 
and x directions were 120 cm and  5 cm, respectively 
(see Fig.~1). The position in the y direction was 
 determined from the differing arrival times of the scintillator 
 light pulses read out by the top and bottom photomultipliers. 
 The y-position resolution was 
 $\sigma \approx 1.7 $ cm. At 325 MeV the geometric 
 acceptance for p + n detection is comparable to that 
for the \ppp \  branch; however, the achievable 
 event detection rate is much smaller because of the
 low neutron detection efficiency.
The neutron pulse height threshold was set as low as practical and
corresponds to 5 MeV electrons for all bars. At this threshold
a 15 cm thick plastic scintillator averages a neutron 
detection efficiency of about 0.17 for the 
neutron energies of this experiment \cite{daehnick92}. 

A thicker neutron detector would be more efficient, 
but along with technical problems it would  
produce a correspondingly poorer time of flight resolution, since 
the length of the available flight path was limited to 1.5 m.
In this experiment an additional reduction of the neutron 
detection efficiency arose because the E and K 
proton detectors are located in front of the neutron hodoscope
and represent a 26 cm thick (polystyrene) absorber for the reaction neutrons.
Resulting neutron losses in this ``absorber'' range from 
{30\%} for the highest energy neutrons to about {90\%} for those at the 
very lowest energies. As a consequence the energy averaged 
effective neutron detection efficiency was reduced to 
a value of about 0.07. 
Since neutron energies are measured and neutron
reaction cross sections are known  \cite{nsigma},
corrections for energy dependent efficiency losses can and have 
been made, but the loss in counting rate seriously limited the 
statistics obtained.

The neutron energy was measured by neutron time of 
flight. In applying this method we use the correlated proton 
trigger from {$\vec{p}\vec{p}\rightarrow pn\pi^+$} in the F detector. 
Since the proton arrival at the F detector is delayed one 
has to use a two-step process: 
First, the trigger time difference (F detector 
time minus hodoscope mean time) is measured.
Next the timing must be corrected for the proton flight 
time to the F detector, since the F detector is triggered 
by the proton  after it has traveled about 30~cm before 
reaching the F detector. This correction is based on the 
measured proton energy and reconstructed track length.  
Neutron flight times (TOF) range from 5 to 12 ns.

The dominant contribution to the TOF resolution 
comes from the 15 cm bar thickness, 
which constitutes 10\% of the flight path and cannot be overcome with 
the available detectors. Smaller contributions come from the intrinsic 
timing resolution of the hodoscope (0.4 ns FWHM) and the 
F detector (0.5 ns FWHM, after amplitude walk correction).
We note that the raw time resolution 
of the F detector is worse than the figure 
quoted above because of trigger walk in the electronics and 
because of the light
loss and travel delay of light from parts of the 
large four-section F detector more distant from the photomultipliers. 
A substantial improvement was achieved by employing a pulse 
height compensation function.
Overall, we see a neutron time of flight resolution with 
$\Delta T/T \approx  0.1 $.
Therefore, the missing mass (MM) peak for $\pi^{+}$ from p+n detection
is not as sharp as for
the corresponding neutron missing mass derived from p+$\pi^{+}$ events.

\section {Analysis}

\subsection {Monte Carlo Simulations}

Our Monte Carlo (MC) simulations of the experiment used the 
event generator GENBOD of the CERN library. The simulation
was used to determine various limiting effects of the apparatus 
and to derive corresponding corrections. The code contained the 
detailed geometry of the detector systems and the  density 
distribution of the gas target. 
In the MC simulation we took into account the loss of energy 
of the charged particles before entering the detectors,
detector resolutions, charged particle multiple
scattering, pion decay in flight, energy-dependent neutron detection
efficiency and the probability of nuclear reactions of the reaction 
neutrons in the E and K detectors. In the MC simulation 
we have used a $pn$ final-state interaction (FSI)
based on the Watson-Migdal theory, 
and the equations were derived following Morton \cite{morton}. 
We found that at the lower energies the FSI has a large 
effect on the overall coincidence acceptance.
The simulation also provided a guide to the expected energy
and angular distribution of the reaction products. 

Pion counting losses caused by the limited detector depth are not 
large enough to be detectable in the shape of spectra; however, 
the Monte Carlo simulation shows 
that they must be considered. 
At 400 MeV the loss for pions is 14\% because this fraction of the 
forward pions is too energetic to stop in the K detector.
Only about 0.2\% of the reaction protons 
penetrate past the K scintillators and are vetoed.
The loss of high energy pions at small lab angles
may create a small distortion of the 400 MeV p+$\pi$ data.
Corrections to the 400 MeV spectra were not made
since they would have to be very model dependent.
We note parenthetically that the 400 MeV data from p+n coincidences
do not have this systematic error, but within statistics they agree 
with overlapping $p+\pi^{+}$ results.
No ``veto'' losses are seen at 375 MeV or below.

The finite size of the individual detector segments produces
some counting losses since two sections have to trigger for acceptable 
events. However, systematic effects for the 
polarization observables are unlikely since the
protons have no strong $\phi$ correlations with the pions. 
The  segmentation used leads to a 
loss of about 7\% in counting statistics for the p+$\pi^{+}$ branch.
There is no such loss for p+n detection.
The charged particle detectors cover polar angles
between $5^{\circ}$ and $40^{\circ}$ in the laboratory frame. 
Hence a large number of pions miss the detector.
The total $p +\pi^{+}$ coincidence acceptance ranges 
from 21\% at 325 MeV to 15\% at 400 MeV.

\vspace{-1.3cm}
\hspace{2.cm}
\begin{figure}[htb]
\epsfxsize=8.cm
\centerline{
\epsfbox{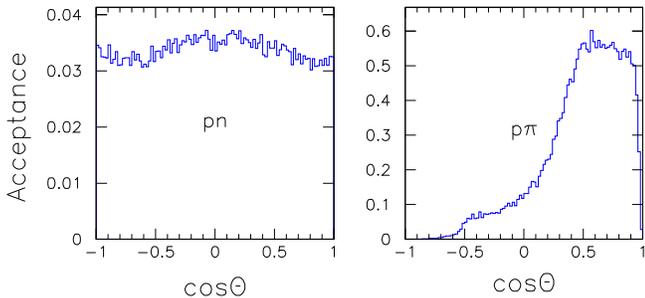}
}
\vspace{-5.6cm}
\caption{Monte Carlo simulation for (a) the p+n and (b) the 
p+$\pi^+$ acceptance, in the center of mass system.
The partial acceptance for pions seen in the $p\pi$
diagram at $\theta_{\pi} \ge 70^{\circ}$
results from the dominating forward boost for low energy pions. The 
cutoff at $cos~\theta_{\pi} \approx - 0.5$ is caused by detector 
thresholds for the lowest ejectile energies.}
\label{fig:accept}
\end{figure}
For p+n detection the MC simulation shows that the acceptance is 
symmetric about $90^{\circ}$ although not quite isotropic.
(See  Fig.~\ref{fig:accept}a.)
Acceptance losses for p+n coincidences attributable to the detector 
geometry alone are of the order of 25\%.  The major cause is the
central hole in the proton detectors.
After all geometric acceptance losses and detector 
inefficiencies for neutron detection are taken into account the 
computed overall detection efficiency for pn 
coincidence events is 3.5\%.
It is seen in Fig.~\ref{fig:accept}a  that the angular variations 
of the coincidence efficiency for the reconstructed pion are small. 
This is so despite the fact that we cannot detect protons at  
angles $\le 5^{\circ}$ and neutrons at angles $\le 2.5^{\circ}$ 
and have reduced coverage by the hodoscope of
some azimuthal angles for large neutron polar angles.
The Monte Carlo acceptance curves for p+n detection suggest 
that within the statistical accuracy of the
experiment the spin-correlation parameters integrated over
$\rm \theta_{\pi}$ and $\rm \phi_{\pi}$ would 
need no significant correction.
Fig.~\ref{fig:accept}b shows the Monte Carlo simulation for
$p\pi^+$  acceptance as a function of $cos~ \theta_{\pi}$.
For $p\pi^+$ coincidences the apparatus acceptance is only
useful for $\theta_{\pi} \leq 70^{\circ}$. 
Therefore, the integrated spin correlation coefficients
will be deduced from the combined sets of the $p+\pi^{+}$ 
and p+n coincidences.

\subsection {Analysis of p+$\pi^+$ coincidences}

The energies of the charged particles are measured 
by the plastic scintillator systems E and K. 
The calculated momenta of the unobserved particles 
strongly depend on the energies of the detected ejectiles, 
so considerable attention was given 
to a careful energy calibration of all detectors.
The complex geometry of the segmented plastic 
detectors required  corrections for light collection 
that primarily were derived 
from the observation of elastically scattered protons.
An xy-position correction factor was applied to account for this
dependence. A second pulse height correction factor was applied
to compensate for a variation of phototube gains with
the orientation of the  magnetic guide field 
for the target polarization. For details see \cite{rinckel}.

 The corrected pulse heights L were
 converted into the deposited energy E using 
%
\begin{equation}
E =   L  + k_1\sqrt{L} + k_0
\label{eq:ecalib}
\end{equation}
The nonlinear term corrects for light quenching in plastic 
scintillators.  $k_0$ and $k_1$ are calibration constants.
L is the sum of the light 
pulse from the E and the K detectors in MeV and is given by
{$L = c_1(E_{light} + c_2K_{light} + c_3)$}.
The constants $c_1,c_2,c_3$ are gain matching constants, 
and $E_{light}$ and $K_{light}$ correspond to the 
observed light pulses in the $E$ and the $K$  detectors 
respectively. The constant $c_3$ corrects for small energy
losses in the material between the E and K detectors. It
is small and set equal to zero when there is no K trigger. 

The total kinetic energy of the charged particle was 
calculated by also taking account of the energy lost
by the charged particle on its way to the E-detector. The
 calibration constants were fine tuned by 
utilizing kinematical relations. We
required that the missing mass centroid was at its predicted 
value and that the angular distribution of the pions from
the simultaneous measurement of the reaction 
$\vec{p}\vec{p}\rightarrow pp\pi^0$ was symmetric in the c.m. system 
about $\theta_{\pi}=90^0$. This symmetry was 
sensitive to the relative size of the calibration 
coefficients. However, the variation of the deduced spin
correlation coefficients under different reasonable combinations
of the calibration constants was small and less than the statistical
errors.
\vspace{-1.cm}
\hspace{-1.cm}
\begin{figure}[htb]
\epsfysize=7.cm
\centerline{
\epsfbox{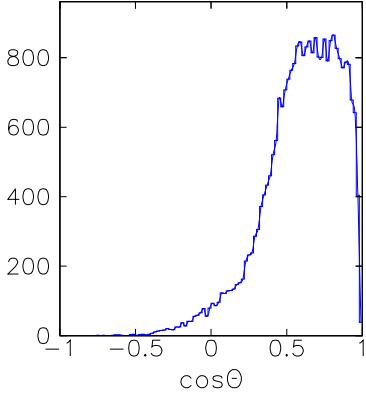}
}
\vspace{-0.5cm}
\caption{ Detected pions at 325 MeV as a function of $\theta_{\pi}$. 
Only events with $cos\,\theta_{\pi} \ge 0.4$ were used for the analysis.}
\label{fig:ppisigma}
\end{figure}
Fig.~\ref{fig:ppisigma} shows the directly observed $\pi^+$ 
differential cross sections plotted against $cos\,\theta_{\pi}$ 
in the c.m. coordinate 
system. We note that there are almost no counts for pion back 
angles $-1 <cos\,\theta_{\pi}< 0$ , as 
expected from the apparatus acceptance (compare Fig.~\ref{fig:accept}b). 

\begin{figure}[htb]
\vspace{-3.5cm}
\epsfysize=14.cm
\centerline{
\epsfbox{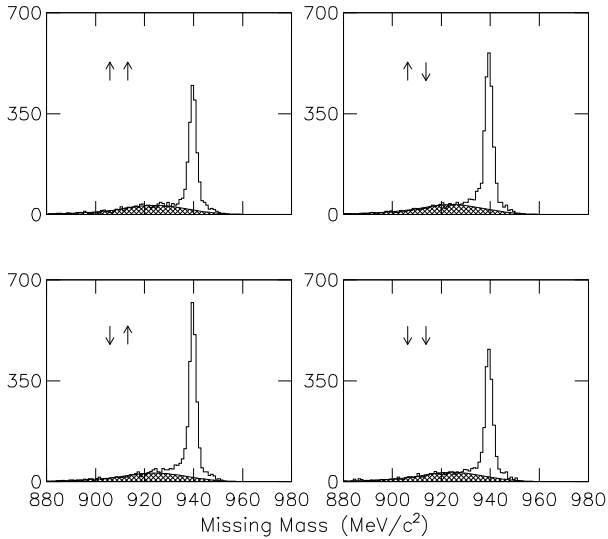}
}
\vspace{-3.cm}
\caption{Distributions of the calculated missing mass $\rm m_x$ for 
$p+\pi^{+}$ detection at 325~MeV bombarding energy, for the four
combinations of vertical beam and target polarization. 
A sharp peak ($\approx \rm 3.5~MeV/c^2$ FWHM) is seen 
at 939.6 $\rm MeV/c^2$, the neutron rest mass. The shaded region 
indicates the background distribution.}
 \label{fig:mm(E)}
\end{figure}
Fig.~{\ref{fig:mm(E)} shows missing mass spectra seen at 325 MeV for 
four combinations of vertical beam and target polarization at equal 
integrated luminosity. The polarization observables are obtained from 
the ratio of ``yields'' for different spin orientations. The yields to
be used are the integrated counts inside the missing mass gates minus 
background.
In order to estimate the error from uncertainties in the background we 
varied the background subtraction by $\pm$25\%. The effect on the final
results was smaller than the statistical error. At 325~MeV the  
off-line resolution
of the neutron missing mass peak was $\rm \sigma = 1.4~ MeV/c^{2}$. 
Even before software cuts and background correction it is apparent from
Fig.~{\ref{fig:mm(E)} that different spin combinations produce 
very different yields. 

It turns out that the decay in flight of pions plays a negligible role
for these data. It will
appreciably affect only the (undetected) backward scattered pions 
as these have much lower lab energies than the forward pions. 
For the final analysis we selected the \ppn events of interest by 
using a gate of 30 MeV or wider over the relevant missing mass peak. 
Gates as narrow as 
10 MeV did not produce systematic changes, neither did they 
measurably reduce background induced errors. However, the narrower 
gates lead to some loss of statistics.

\subsection {Analysis of p+n coincidences}

This detection channel has the advantage that the acceptance 
for the detection of p+n coincidences has little 
angular variation. So  $\theta_{\pi}$ 
and $\phi$-dependent acceptance corrections generally
can be ignored. Therefore the p+n coincidences
importantly complement the p+$\pi^{+}$ channel. 
Reliance on the p+n angular distributions at large angles
 leads to larger statistical
 error bars relative to p+$\pi^{+}$ (forward) region. 
However, the combination of the two detection modes 
provides data for the full angular range and so keeps 
the integrated spin correlation coefficients model independent.

In the p+n analysis we first analyzed
only those events where all three reaction particles ($p,n$ and
$\pi^+$) were detected (the triple coincidence). Next we evaluated
the case where the pions missed the E detector, but a proton and a 
neutron were detected (double coincidence).
The energy of reaction protons was determined using 
the calibration constants described above. 
The  energy of the neutrons was determined by measuring 
their time of flight (TOF) to the hodoscope. 
The MC simulation showed that although F was always 
triggered by protons for a p+n double coincidence, 
in the case of a $pn\pi^{+}$ triple coincidence
it was triggered by the faster pions. Therefore, depending on 
the event class, we corrected the 
neutron TOF by adding the time it takes either for the coincident 
proton or the pion to reach the F detector. 
A calculated offset was added to the timing signal of each hodoscope bar
in order to make the timing information independent of the bar 
electronics. This correction was obtained by calibrating the timing 
circuits with elastic proton scattering.

For $\theta_{\pi,lab} \le 40^{\circ}$ we observe  
$pn\pi^{+}$ triple coincidences, which are practically free of 
background. 
(The absence of accepted events from the $N_2$ gas target 
showed that the triple $pn{\pi^+}$ hardware
coincidence under standard software conditions eliminates all
background from the target wall and target impurities.)
These events proved very valuable in assessing the correct 
shape of the  missing mass peaks in p+n and p+$\pi$ events. If a missing 
mass spectrum for triple coincidences is calculated based on the pion and 
proton momenta the (neutron) missing mass spectrum shows a very sharp 
peak as in Fig.~4, but there is no ``background tail'' at all. The 
triple coincidence spectrum confirms the background subtraction shown 
in Figs.~\ref{fig:nmm} and \ref{fig:mm(E)}.

If the same triple coincidence events are used to calculate 
the (pion) missing mass by 
using the proton and neutron momenta (i.e., ignoring the 
simultaneously known pion momenta) we obtain the spectrum shown in 
Fig.~\ref{fig:triplemm}. This spectrum can be used as a standard  
for the missing mass that p+n (double coincidence) events would have 
in the absence of background.
%
%
\vspace{-.75cm}
\begin{figure}[htb]
\epsfysize=7.cm
\centerline{
\epsfbox{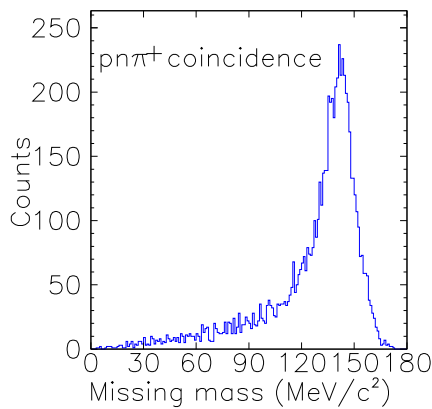}
}
\vspace{-0.5cm}
\caption{Missing mass spectrum for $pn\pi^{+}$ triple coincidences, 
based on the measured neutron and proton energies. This spectrum 
contains no background from any competing reaction. The missing mass 
tail here is a consequence of some inaccurately measured 
neutron momenta. See text.}
\label{fig:triplemm}
\end{figure}
The MM distribution peaks at the true pion mass of 
139.6~MeV, but there  also is a ``tail'' over a
wide range of the missing mass spectrum, which is not background 
related.
 We conclude that the counts in the MM tail of Fig.~\ref{fig:triplemm} 
represent genuine 
$pn\pi^{+}$ events from the Hydrogen target, albeit events 
with poorly determined neutron momenta. 
We estimate that up to 20\% of the p+n coincidences contain 
neutron observables that are 
distorted by interactions of neutrons with the 
K or E detectors. E.g.,  neutrons can undergo 
small angle elastic and inelastic scattering, but still 
reach the hodoscope. This would lead to incorrect
readings for polar and azimuthal neutron angles and hence to 
an incorrect missing mass calculation.

Such events with poorly determined missing masses 
were excluded from further analysis. 
For all p+n events we reduce genuine background and 
avoid analyzing measurably distorted p+n events 
by using a missing mass gate from 100 to 160 MeV.
%
\vspace{-1.cm}
\hspace{-1.cm}
\begin{figure}[htb]
\epsfysize=7.cm
\centerline{
\epsfbox{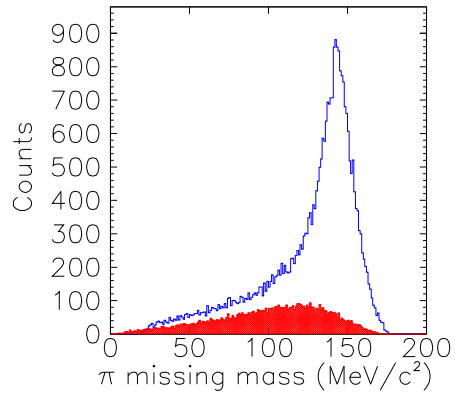}
}
\vspace{-1cm}
\caption{The $\pi^+$ missing mass ($m_{x}$) spectrum at 375 MeV
calculated from the measured neutron and proton momenta. The  
deduced background is shown by the lower distribution. For the 
analysis events with $100 \le m_{x} \le 160$ were accepted.}
\label{fig:mmpi+}
\end{figure}
Using the triple coincidence MM spectrum as a standard,
the background under the missing mass peak for two-particle p+n
coincidences was deduced by adding a fraction of the measured unstructured $N_2$ 
background continuum to the ``standard'' MM spectrum until the observed 
p+n MM spectrum shape was reproduced.
The tail in the latter is flatter and more pronounced because 
of actual background contributions. 
To estimate the error in this procedure we varied the match
until it became unrealistic ($\pm15\%$).
A typical missing mass spectrum for p+n (double coincidence) detection is 
shown in Fig.~\ref{fig:mmpi+}.
In the final result the uncertainty from this background subtraction 
was about half as large as the statistical error. 
%
\vspace{-1.2cm}
\begin{figure}
\epsfxsize=7.5cm
\centerline{
\epsfbox{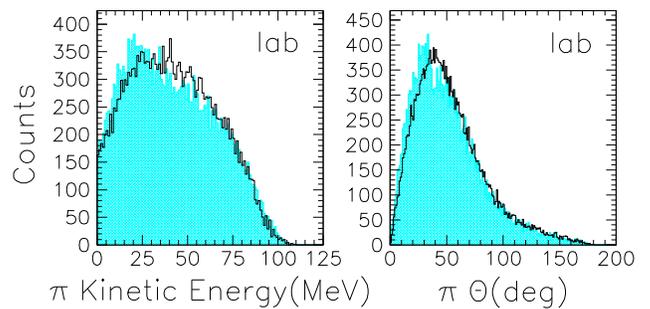}
}
\vspace{-5.cm}
\caption{Energy and angular distributions 
for p+n coincidences at 375 MeV, compared with Monte Carlo projections
(solid lines) in the laboratory system.  
Agreement is expected for the kinematic limits. However,
the distributions may differ because the  $l_{\pi}=0$  
assumption for the MC simulation is an oversimplification 
at and above 325 MeV.}
\label{fig:pnkeang}
\end{figure}
At 375~MeV the resolution of the pion MM peak was 
$\rm \sigma = 9~MeV/c^{2}$. Pion angular and energy 
distributions from p+n detection were computed 
using only events inside this missing mass gate. 
Some resulting distributions are compared with Monte Carlo 
projections for the laboratory coordinate system 
in Fig.~\ref{fig:pnkeang}.
\onecolumn
\narrowtext
The end points of these distributions agree well with the
kinematics of the experiment as they must. 
The solid curves represent pure $\l_{\pi}=0$ MC 
calculations. 
Although $\l_{\pi}=0$ makes the major
contribution, this MC assumption produces oversimplified 
energy and angular distributions. 
Nevertheless, the simulated distributions agree reasonably 
well with the data. 

Fig.~\ref{fig:pnsigma} shows the deduced pion angular distribution 
in the center of mass system. The reconstructed $\pi^+$ distribution 
is plotted against $cos\,\theta_{\pi}$ in the center of mass 
(corrected for background and for the slightly
non-uniform acceptance shown in Fig.~\ref{fig:accept}a).
As expected, it is non-isotropic and symmetric
about $\theta_{\pi}=90^o$ within statistical errors.
\vspace{-1.cm}
\begin{figure}[htb]
\epsfysize=7.0cm
\centerline{
\epsfbox{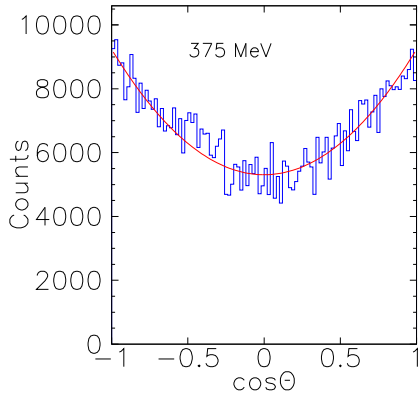}
}
\vspace{-0.7cm}
\caption{The relative $\pi^{+}$ production cross section 
$\sigma_{0}(cos~\theta_{\pi})$ at 375 MeV as deduced from p+n 
coincidences.}
\label{fig:pnsigma}
\end{figure}
\vspace{-0.2cm}
\section {Polarization Observables}

\subsection{Formalism for spin correlation coefficients}
The  meaning of the symbols $A_{ij}$ used for polarization 
observables is defined by Eq.~\ref{sigma_cart}.
In terms of the ``Cartesian  
polarization observables'' the spin--dependent cross section 
is written as
\widetext
\begin{equation} \label{sigma_cart}
\sigma(\xi,\vec{P},\vec{Q})\ =\ 
       \sigma_0(\xi)\cdot\left[ 1 + \sum_i P_i A_{i0}(\xi) +
       \sum_j Q_j A_{0j}(\xi) + 
       \sum_{i,j} P_iQ_j A_{ij}(\xi)\right]\ ,
\end{equation}
\narrowtext
\noindent
where $\xi$ stands for the pion coordinates 
$\rm \theta_{\pi},\varphi_{\pi}$, the 
energy defining pion momentum $p_{\pi}$, and the
proton coordinates $\rm \theta_{p}$ and $\rm \varphi_{p}$.
The unpolarized cross section is $\sigma_0(\xi)$, and
the polarization of the beam and the target is denoted 
by the  vectors $\vec{P} = (P_x, P_y, P_z)$
and $\vec{Q} = (Q_x, Q_y, Q_z)$. The subscripts $i$ and $j$ 
stand for $x$, $y$ or $z$ and the sums extend  
over all  possibilities. The resulting 15 polarization observables
include the beam analyzing powers A$_{i0}$, the target analyzing
powers A$_{0j}$ and the spin correlation coefficients A$_{ij}$.

The partial wave analysis for $\vec{p}\vec{p}\rightarrow pn\pi^+$
is similar to that for $\vec{p}\vec{p}\rightarrow pp\pi^0$
in terms of transition amplitudes. However, the \ppn \ transitions
have isoscaler as well as isovector components.
The different isospins in $\vec{p}\vec{p}\rightarrow pn\pi^+$} 
modify the selection rules for the reaction and
lead to polarization observables that 
are different. The general relations between reaction 
amplitudes and angular
distributions, however, remain almost identical.
The applicable partial wave formalism has been 
discussed in detail in \cite{meyer01}. 
We  use the same notation as in \cite{meyer01} and 
reiterate some relevant definitions and theoretical 
relations below.
Several names are in use for polarization observables. 
Their meaning is as defined below:
\begin{mathletters}
\label{A_SDX}
\begin{eqnarray} 
A_{\Sigma}(\xi)\ \equiv\ A_{xx}(\xi) + A_{yy}(\xi)\ \\
A_{\Delta}(\xi)\ \equiv\ A_{xx}(\xi) - A_{yy}(\xi)\ \\
  A_{\Xi}(\xi)\ \equiv\ A_{xy}(\xi) - A_{yx}(\xi)\ 
\end{eqnarray}
\end{mathletters}
For identical particles in the entrance channel there are seven 
independent polarization observables:
\mediumtext
\begin{equation} \label{obs_cart}
  A_{y0}(\xi),  \ A_{\Sigma}(\xi), \  A_{zz}(\xi),
  \ A_{xz}(\xi),\ A_{\Delta}(\xi),\ A_{\Xi}(\xi)\ ,\ A_{z0}(\xi).
\end{equation}
\narrowtext
This paper addresses the first five observables of this set. The remaining 
two,  $A_{\Xi}(\xi)$ and $A_{z0}(\xi)$, can be non-zero only for 
non-coplanar final states. In the following we will integrate over 
the angles of the nucleon, and thus these two observables vanish
if parity is conserved.

It is common to display the bombarding energy dependence of the observables
in terms of the dimensionless parameter eta (${\eta}$), which is defined
as
\begin{equation} 
\label{def_eta}
\eta\ =\ p_{\pi,max}/m_{\pi^{+}}\ ,
\end{equation}
 The term ``near threshold'' is meant to include
the energy region with  $\eta < 1$, i.e., below 400 MeV.
Setting $c=\hbar=1$, the maximum value of the $\pi^{+}$ 
momentum is found from:
%
\widetext
\begin{equation} \label{def_qmax}
p_{\pi,max}\ =\ \frac{1}{2\sqrt{s}}\sqrt{[s-(m_{n}+ m_p+m_{\pi^{+}})^2]
                                     [s-(m_{n}+m_p - m_{\pi^{+}})^2)]}\ ,
\end{equation}
\narrowtext
\noindent
where $\sqrt{s}$ is the total center-of-mass energy, and
$m_p$,  $m_n$, and $m_{\pi^{+}}$ are
the masses of the proton, the neutron, and the pion, respectively. 
(We explicitly labeled the pion as $\pi^{+}$ to emphasize that 
the $\pi^{+}$ and $\pi^{0}$ mass difference matters here.)

Below we quote some useful relations between integrated spin correlation 
coefficients and some directly observable spin dependent cross
sections. For two colliding spin 1/2 particles, one can
define  three  total cross sections, two of which depend on the
spin.  The total cross sections are related to the observables
above by
\begin{mathletters}
\begin{eqnarray}
\sigma_{tot} & = & \int\sigma_0(\xi)d\Omega_p d\Omega_{\pi} d p_{\pi} \\
\Delta\sigma_T & = & 
  -\int\sigma_0(\xi)A_{\Sigma}(\xi)d\Omega_p d\Omega_{\pi} d p_{\pi} \\
\Delta\sigma_L\ & = & 
  -2 \int\sigma_0(\xi)A_{zz}(\xi)d\Omega_p d\Omega_{\pi} d p_{\pi}
\end{eqnarray}
\end{mathletters}
\noindent
Here $d\Omega = d\cos\theta \, d\varphi$, and the integration extends
over $0 \le \theta \le \pi$, and all pion momenta.  
$\Delta\sigma_L/\sigma_{tot}$ and
$\Delta\sigma_T/\sigma_{tot}$ can have values between $-2$ and +2.

The integrated spin correlation coefficients  are defined as:
\begin{mathletters}
\begin{eqnarray}
\overline{A_{\Sigma}} & = & 
  [\int\sigma_0(\xi)A_{\Sigma}(\xi)d\Omega_p d\Omega_{\pi} d p_{\pi}]/ \sigma_{tot}\\
\overline{A_{zz}} & = & 
 [\int\sigma_0(\xi)A_{zz}(\xi)d\Omega_p d\Omega_{\pi} d p_{\pi}]/ \sigma_{tot}\\
\overline{A_{\Delta}} & = & 
  [\int\sigma_0(\theta_{\pi})A_{\Delta}(\theta_{\pi})sin\theta_{\pi} d\theta_{\pi}]/ \sigma_{tot}\\
\overline{A_{xz}} & = & 
 [\int\sigma_0(\theta_{\pi})A_{xz}(\theta_{\pi})sin\theta_{\pi} d\theta_{\pi}]/ \sigma_{tot}\\
\overline{A_{y0}} & = & 
 [\int\sigma_0(\theta_{\pi})A_{y0}(\theta_{\pi})sin\theta_{\pi} d\theta_{\pi}]/ \sigma_{tot}
\end{eqnarray}
\end{mathletters}
\noindent
We note that $\overline{A_{\Sigma}}$ and $\overline{A_{\Delta}}$ differ by a 
scale factor from $\Delta\sigma_T$ and $ \Delta\sigma_L $. 
These quantities in principle can be measured directly, although in 
this study they are derived from integration over 
$A_{\Sigma}(cos \theta_{\pi})$ and $A_{zz}(cos \theta_{\pi})$. The remaining three
integrals must be defined differently. Here the spin correlations $A_{ij}$ are taken 
at $\phi_{\pi}=0$. (They cannot be integrated over the variable 
$\phi_{\pi}$ since they would vanish, as will be seen below).\\

Based on the dominance of Ss, Sp, Ps, and Pp transitions, general 
symmetries and spin coupling rules \cite{meyer01}, the cross 
sections and spin correlation coefficients must have the general form:
\widetext
\begin{mathletters}
\label{mast_eq}
\begin{eqnarray}
\sigma_0(\xi) & = & a_{00} +
                    b_{00} \ccq +
                   c_{0}\ccp \nonumber\\
              &   & +d_{0}\ccq\ccp
                    +e_{0}\stp\stq\cos\Delta\varphi \nonumber\\
              &   & + f_{0}\ssp\ssq\cos 2\Delta\varphi \\[1.2ex]
\sigma_0(\xi)A_{y0}(\xi) & = & [{\{a_{y0}+b_{y0}\ccp\ \}\sinq +
                                \{c_{y0}+d_{y0}}\ccp\} \stq]
                               \cos\varphi_{\pi} \nonumber\\
                         &   & + [e_{y0} +  f_{y0}\cosq +
                                  g_{y0}\ccq]\stp\cos\varphi_p \nonumber\\
                         &   & + [h_{y0}\sinq + i_{y0}\stq]\ssp
                                 \cos(2\varphi_p - \varphi_{\pi}) \nonumber\\
                         &   & + j_{y0}\stp\ssq
                               \cos(2\varphi_{\pi} - \varphi_p) \\[1.2ex]
\sigma_0(\xi)A_{\Sigma}(\xi) & = & a_{\Sigma} + 
                                   b_{\Sigma}\ccq +
                                   c_{\Sigma}\ccp \nonumber\\
                             &   & + d_{\Sigma}\ccp\ccq +
                                   e_{\Sigma}\stp\stq\cos\Delta\varphi 
                                   \nonumber\\
                             &   & + f_{\Sigma}\ssp\ssq\cos 
                                   2\Delta\varphi \\[1.2ex]
\sigma_0(\xi)A_{zz}(\xi) & = & a_{zz} + 
                               b_{zz}\ccq +
                               c_{zz}\ccp \nonumber\\
                         &   & + d_{zz}\ccp\ccq +
                               e_{zz}\stp\stq\cos\Delta\varphi \nonumber\\
                         &   & + f_{zz}\ssp\ssq\cos 
                               2\Delta\varphi \\[1.2ex]
\sigma_0(\xi)A_{\Delta}(\xi) & = & [a_{\Delta} + b_{\Delta} \ccp]\ssq
                                   \cos 2\varphi_{\pi} \nonumber\\
                             &   & [c_{\Delta}  + d_{\Delta} \ccq]\ssp
                                   \cos 2\varphi_p \nonumber\\
                             &   & e_{\Delta} \stp\stq
                                   \cos(\varphi_p + \varphi_{\pi}) \\[1.2ex]
\sigma_0(\xi)A_{xz}(\xi) & = & [\{a_{xz} + b_{xz}\ccp\}\sinq +
                                \{c_{xz} + d_{xz}\ccp\}\stq]
                               \cos\varphi_{\pi} \nonumber\\
                         &   & + [e_{xz}  + f_{xz}\cosq +
                                  g_{xz}\ccq]\stp\cos\varphi_p \nonumber\\
                         &   & + [h_{xz}\sinq + i_{xz}\stq]\ssp
                                 \cos(2\varphi_p - \varphi_{\pi}) \nonumber\\ 
                         &   & + j_{xz}\stp\ssq
                                 \cos(2\varphi_{\pi} - \varphi_p)
\label{mast_eq_g}
%
\end{eqnarray}
\end{mathletters}
\twocolumn
\noindent
Here we have used the abbreviation 
$\Delta\varphi \equiv \varphi_p - \varphi_{\pi}$.
The Eqs.~\ref{mast_eq} explicitly depend on the four angles $\theta_p$, 
$\varphi_p$, $\theta_{\pi}$ and $\varphi_{\pi}$. The energy dependent 
parameter $p_{\pi}$ is contained in the coefficients.  
Statistics in this experiment are not sufficient to present 
double or higher differential cross sections. 
Therefore, we integrate over 
the angles of the proton and use energy and momentum conservation 
to eliminate all angles except $\theta_{\pi}$ and $\phi_{\pi}$.
This leads to a set of much simpler equations:

\begin{mathletters}
\label{fit_eq}
\begin{eqnarray}
\sigma_0(\zeta) & = & a_{00} + b_{00} \ccq  \\[1.2ex]
\sigma_0(\zeta)A_{y0}(\zeta) & = & [ {a_{y0}} \sinq + {c_{y0}} \stq] 
\cos\varphi_{\pi} \\[1.2ex]		       
\sigma_0(\zeta)A_{\Sigma}(\zeta) & = & a_{\Sigma} + {b_{\Sigma}} \ccq  \\[1.2ex]
\sigma_0(\zeta)A_{zz}(\zeta) & = & a_{zz} + {b_{zz}}\ccq  \\ [1.2ex]
\sigma_0(\zeta)A_{\Delta}(\zeta) & = & {a_{\Delta}}\ssq
                                   \cos 2\varphi_{\pi} \\ [1.2ex]
\label{mast_eq_h}
   \sigma_0(\zeta)A_{xz}(\zeta) & = & [{{a_{xz}} \sinq + {c_{xz}}} \stq] \cos 
   \varphi_{\pi}  \\
\nonumber
\end{eqnarray}
\end{mathletters}
The symbol $\zeta$ now represents the reduced set of variables 
\{$p_{\pi},\theta_{\pi},\phi_{\pi}$\}.
These equations display a simple and characteristic $\phi_{\pi}$ 
dependence of the different polarization observables and show the
expected $\theta_{\pi}$ dependence.
The coefficients $a_{\nu}, b_{\nu}, \ldots$ for the set \ref{fit_eq} 
correspond to those in Eqs.~ \ref{mast_eq}.  
They are obtained by one- or two-parameter 
fits to the observed angular distributions, separately for each 
observable.

\subsection{Extraction of polarization observables}

The data analysis, as described in the previous sections, identifies the 
reaction particles, assesses the background for each spectrum and 
calculates the kinematic variables and spin dependent cross sections 
of the reaction products. It produces event files which contain 
kinematically complete information
for all detected reaction particles. For each beam energy 
there are 12 such event files, one
for each combination of beam and target spin. 
These yields are first corrected for the 
beam luminosity, which can vary for beam ``spin up'' 
and ``spin down'' subcycles, and for the background measurement. 
The background correction was made for each selected $\theta_{\pi}$ 
angle bin individually.
The  ratios $R_{i}$ of yields for different
spin combinations, integrated over a chosen $\theta_{\pi}$ range, 
are then analyzed as a function of $\phi_{\pi}$, because the allowed 
$\phi_{\pi}$ dependence can be predicted from spin coupling rules 
\cite{knutsen}.
For this energy range, only final states with  pn or pion
angular momentum of 0 and 1 are expected to be significant.
In a previous measurement of $pp\rightarrow d\pi^{+}$ at 400 MeV 
it was found that any $l_{\pi}=2$ contribution is very small 
\cite{przewoski00}.
This allows us to consider only transitions to Ss, Sp, Ps and Pp 
final states in the analysis. Sd and Ds transitions would 
affect the energy dependence of the coefficients only and so are 
very difficult to separate from Pp transitions
\cite{meyer01}. They will be ignored in this analysis. We then have 
explicit predictions for the expected $\theta$ and $\phi$ dependences 
from eq. \ref{fit_eq}.

The combination of p+$\pi^{+}$ and p+n measurements provides 
model-independent values for the polarization observables
for all polar and azimuthal angles of the pion.
 The low neutron detection efficiency
and the resulting low statistical accuracy of the p+n data
make it advisable to display the combined data using some theoretical 
guidance. As shown below, $A_{\Sigma}(\theta_{\pi}), A_{\Delta}(\theta_{\pi})$,
and $A_{zz}(\theta_{\pi})$ must be symmetric about 
$\theta_{\pi}=90^{\circ}$ for the transitions considered.
So a good analysis in terms of the pion coordinates does not 
require the (redundant) data at large polar angles.
This simplification and the fact that all published theoretical 
predictions have been presented in terms of the pion coordinates
makes these coordinates our preferred system for the analysis.

The microscopic relations between the coefficients and the
transition amplitudes can be derived from the  partial-wave expansion  
described in the appendix of \cite{meyer01}, but they are  
complicated. Moreover, the number of individual \ppn amplitudes 
contributing above 350 MeV has become too large (19 rather than
the 12 for \ppp), since isospin 1 and 0 are present in the final 
state. 
They could not be deduced individually from the \ppn data available.
%
\begin{figure}[htb]
\vspace{-3.cm}
\hspace{-1.cm}
\epsfysize=12.cm
\centerline{
\epsfbox{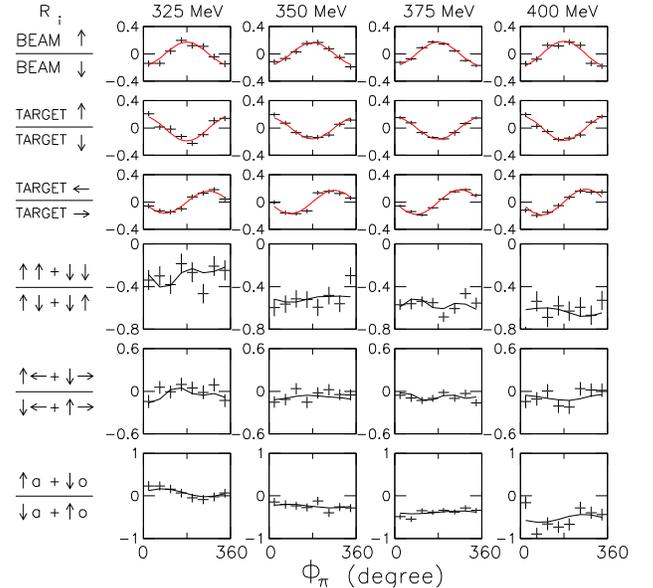}
}
\vspace{-1.cm}
\caption{The yield related ratios $(R_i-1)/(R_i+1)$ as a function of the 
pion azimuthal angle $\phi_{\pi}$ for data integrated over all other 
coordinates. The specific beam and target spin combinations selected 
are listed on the left. For multiple arrows the orientation of the
beam spin is shown first. Longitudinal polarization is indicated by 
the symbols o (opposite) and a (along beam direction).
The solid curves represent a least-square fit using the expected
theoretical $\rm \phi_{\pi}$ dependence.} 
\label{fig:Y(phi)}
\end{figure}
   When calculating the value of a polarization observable from
Eqs.~\ref{mast_eq} or \ref{fit_eq}, one evaluates the ratio 
$A_{ij}(\xi) = \sigma_0(\xi)A_{ij}(\xi)/\sigma_0(\xi)$, so the overall 
normalization of all terms in these equations cancels.
As seen from Eqs.~\ref{fit_eq}  
the yield ratios $R_{i}(\phi_{\pi})$ could either be constant or
have a $\phi_{\pi}$ or 2$\phi_{\pi}$ dependence.
This is borne out by the data
(compare Fig.~\ref{fig:Y(phi)}).

The polarization observables were deduced by 
evaluating the observed $\phi_{\pi}$ dependences of the ratios $R_{i}$ 
for selected beam and target spin combinations. 
This evaluation is complex when longitudinal as well as
transverse beam polarizations are present at the same time. 
Therefore, the devolution process uses the computerized fitting 
routine BMW \cite{balewski}, which was written for this purpose.

Fig.~\ref{fig:Y(phi)} shows the $\phi_{\pi}$ dependence of six spin 
dependent yield ratios. The data for 
the beam (first arrow) and target spin combinations indicated
have been integrated 
over all coordinates other than the coordinate ${\phi}_{\pi}$.
The curves are fits using one to three components of the Eqs. \ref{fit_eq}.
The first three rows present different ways to extract the analyzing 
power $A_{y}(\phi_{\pi})$. 
The lower three rows contain information on $A_{\Sigma} = A_{xx}+A_{yy}$,
$A_{\Delta} = A_{xx}-A_{yy}$, and 
contributions from $A_{zz}$ and $A_{xz}$.
For some ratios the statistics are marginal, and one
cannot exclude the potential presence of 
$\,l$ components higher than included in the analysis, but 
the $\phi_{\pi}$ dependent fits show that the inclusion of 
Ss, Sp, Ps, and Pp transitions
is sufficient to reproduce the data within experimental errors.

For the simultaneous detection of neutrons and protons, 
our data sample the full range for $\theta_{\pi}$, 
although with low statistics. 
We combine the p+$\pi$ and p+n data sets 
to obtain optimal spin correlation coefficients for the 
full angular region. 
We avoid difficulties generated by the non-uniform 
detector acceptances in $\theta_{\pi}$ by evaluating the p+n and 
 $p+\pi$ relations $A_{ij}(cos~\theta_{\pi})$,
which are ratios of cross sections at a given angle.
(Our detection efficiency does not depend 
on spin, and the detector acceptances 
cancel out for $A_{ij}(cos~\theta_{\pi})$.)
The combined  p+n and p+$\pi^{+}$  
sets yield complete angular distributions
with their best statistics at forward angles.
The unpolarized angular distribution $\sigma_{0}(\theta_{\pi})$ 
was obtained to sufficient accuracy from the p+n branch. 
The angular distributions can now be integrated.  
To best account for experimental errors we have chosen to 
integrate Eqs.~\ref{fit_eq} directly after 
the fitting coefficients are deduced. 

Some of the polarization ratios measured are not
independent, as the first three rows in Fig.~12 show.
The reaction has additional redundancies. If parity is conserved and if we
have identical particles in the entrance channel this 
redundancy can give us back-angle information for $A_{y}$ even
though our detectors only cover forward polar angles for the
direct detection of pions.
The correlation we use repeatedly is
\begin{equation}
\label{equivalent} 
A_{ij}(\theta_{p},\phi_{p},\theta_{\pi},\phi_{\pi}) \\
= A_{ji}(\pi-\theta_{p},\phi_{p}+\pi,\pi-\theta_{\pi},\phi_{\pi}+\pi)
\end{equation}

This relation holds for $\rm i \neq j$ and also 
for i=j. E.g., since both $A_{xz}$ and $A_{zx}$
 are measured at forward angles, 
we will obtain the back angle information
for  $A_{xz}$ from the $A_{zx}$ measurement at forward angles.
The polarization observables $A_{xz}$ and $A_{zx}$ are not 
symmetric about $\theta_{\pi}=90^{\circ}$, so this redundancy 
becomes very useful.

\section{Results}
\subsection{Polarization observables}

It follows from Eq.~\ref{equivalent} that
the observables $A_{\Sigma}$, $A_{\Delta}$, and $A_{zz}$
are symmetric about $\theta_{\pi}=90^{\circ}$ 
($cos\,\theta_{\pi}=0$). Within statistical errors the
experimental data agree with this expectation.
In Fig.~13 we have reduced the scatter 
from the low statistics of the p+n coincidences 
by combining the corresponding data for forward and back 
polar angles.
The data for $cos\,\theta_{\pi} \ge 0.5$ are dominantly determined by
events from $p+\pi$ coincidences. In agreement with theoretical
expectations there is only a slow dependence on the polar angle,
so the lack of good statistics near $\theta=90^{\circ}$
does not impede comparison with theory or the extraction
of good values for the integrated polarization observables.

Fig.~\ref{fig:fig14} shows results for $A_{y}(\theta_{\pi})$ and 
$A_{xz}(\theta_{\pi})$ for the full angular range so 
potential asymmetries can be seen. 
The statistically most accurate data were obtained for 375 MeV.
Here and at 400 MeV the J\"ulich model is at odds with the data. The
fit with Eq.~\ref{fit_eq} (solid lines) does much better.
Still, a close inspection of the fits shows some small, but statistically 
significant differences between the partial wave curve and the data 
at very small and very large angles. 
We also see from Table I that the $\chi^{2}$ value for the $A_{y}$ fit has  
become large.
The $A_{y}$ data suggest that  
higher partial waves enter at 375 MeV, but the
experimental uncertainties discourage the extraction of 
relatively small contributions.

The fits obtained with Eq.~\ref{fit_eq} are good (i.e., 
$\chi^{2} \approx 1$ for all curves except for $A_{y}$ at 
375 and 400 MeV). Therefore Eqs.~\ref{fit_eq} 
together with the coefficients of Table II can
be used to represent the new data. The coefficients in these equations
are bilinear sums of the reaction amplitudes. Their experimental
values are given in Table II.
This set is also used to obtain the integrated spin correlation 
coefficients.
Integration of the angular distributions shown above produces the spin
correlation coefficients in Cartesian coordinates.
These coefficients
were the original objective of this experiment. They are now 
known with good statistical accuracy and are given in Table III.
A comparison of these integrated polarization observables
as a function of beam energy with predictions of the J\"ulich 
model is shown in Fig.~\ref{fig:Aij(E)15}. 

For completeness we note here that our attempt to extract non-coplanar
angular distributions for $A_{z0}$ produced results
(not shown) that were consistent with zero. This is different 
from the \ppp results, but has no obvious explanation.
\onecolumn
\narrowtext 
\widetext 
\vspace{-0.2cm}
\begin{figure}[htb]
\epsfysize=11.4cm
\centerline{
\epsfbox{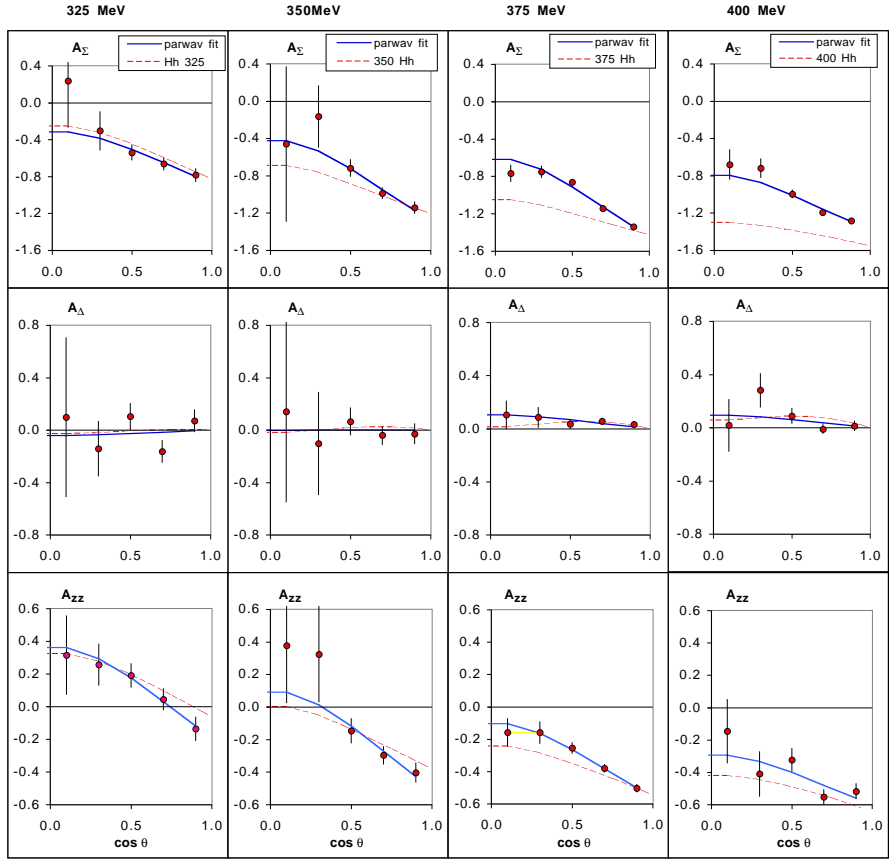}
}
\vspace{0.2cm}
\caption{$\theta_{\pi}$ dependence for the polarization observables
 $ A_{\Sigma} \equiv  A_{xx}+A_{yy}$, 
$ A_{\Delta}\equiv A_{xx}-A_{yy}$, and $A_{zz}$, 
in cartesian units. Data for the 
range $0.5 <~ cos~\theta~<1$ are primarily determined by the  
p+$\pi$ coincidences. The remaining points come from the p+n 
coincidences. The error bars include all random errors as well as 
estimated  uncertainties from background subtraction.
The dashed lines are J\"ulich model predictions by Hanhart et al.[22].
The solid lines show fits with the Eqs.~\ref{fit_eq}.}
\label{fig:fig13}
\end{figure}
%
\vspace{-0.75cm}
\hspace{-3.cm}
\begin{figure}[htb]
\epsfysize=8.2cm
\centerline{
\epsfbox{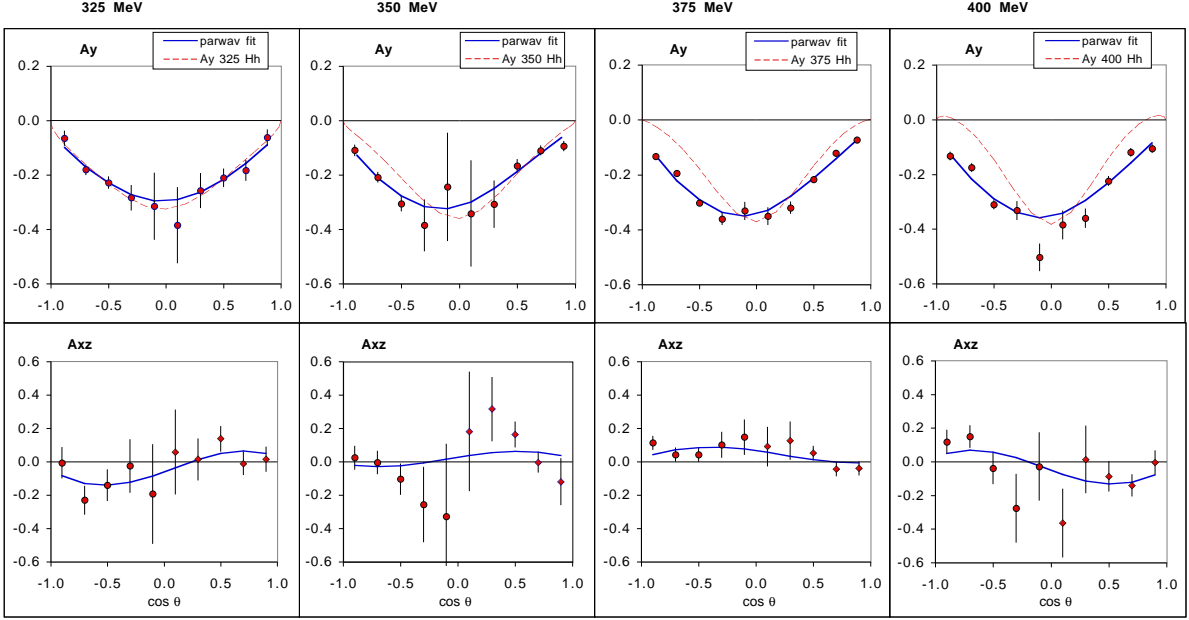}
}
\vspace{0.2cm}
\caption{$\theta_{\pi}$ dependence for the polarization observables $A_{y}$ 
and $A_{xz}$ in cartesian units. The dashed lines are J\"ulich predictions.
The solid lines show fits to the data using the Eqs.~\ref{fit_eq}.}
\label{fig:fig14}
\end{figure}
\narrowtext
\twocolumn
It is clear from  Figures \ref{fig:fig13}, \ref{fig:fig14}, 
and \ref{fig:Aij(E)15} that the  
distributions based on the J\"ulich 
model are in good agreement with the data at 325 MeV.
However, above $\eta=0.7$ they 
produce ever larger $\chi^{2}$ values when compared to the data. 
These disagreements become striking for $A_{\Sigma}$ and $A_{y}$.
The failures are most visible for $A_{y}$, 
an observable sensitive to admixtures of higher partial waves.
(More serious disagreements with this model have been seen for the 
isovector production in \ppp \cite{meyer01}. 
However, as discussed below,  
in this energy region isovector terms contribute less than 10\% 
to the \ppn cross section. The observed differences
in \ppn grow well beyond this level.)
 
Our partial wave analysis, which includes  
Ss, Sp, Ps, and Pp transitions, generally provides  
fits to the measured angular distributions 
with $\chi^{2}$ (per degree of freedom) values near 1.
The exceptions are $A_{y}$ at 375 and at 400 MeV, where the 
cross sections are largest and the statistics 
are good. Some $\chi^{2} $ values as large as 3.9 
are found if only statistical errors are considered. 

The values for the product P*Q are known to good precision (see Table I),
but errors for the beam (P) or target (Q) polarization individually 
are not negligible at the lower energies. Changes in P and Q affect 
only the analyzing powers $A_{y}(\theta)$. They could
reduce or increase the asymmetry of the  angular distributions. Typically, 
the uncertainties in P are smaller than the statistical errors.

\subsection{Discussion and comparison with other work}
%
The statistical and fitting errors listed in Tables II and III 
include all known and estimated random errors.
As explained above, all angles were measured simultaneously,
and systematic normalization errors for $A_{ij}$ are unlikely.
Based on the detector design and redundant measurements 
we expect that all systematic errors have remained small.
In the center region (cos~$\theta \approx 0$) the angular 
distributions show large statistical errors. However, 
these data points do not materially affect the  
partial wave fits or the integrals.
 We note that our initial results reported in \cite{saha99} were
subject to some model dependence that is absent here. Nevertheless, they
are consistent with the final results presented here. 
Noticeable asymmetries around $90^{\circ}$ have been seen for $A_y$ 
above 350 MeV. In the framework of our partial wave analysis this 
asymmetry must be produced by Pp transitions. (At higher energies 
such asymmetries can also be produced by Ds and Sd transitions.)
With the possible exception of the analyzing powers 
$A_y(\theta)$ at 375 and 400 MeV, the pion production data are   
well represented by the partial wave predictions based 
on the the assumption of Ss, Ps, Sp, and Pp transitions.
The J\"ulich model predictions and the data agree for $A_{\Delta}$. 
However, we see serious disagreements for 
$A_{\Sigma}$ and $A_{y}$ as the beam energy increases. Ref. 
\cite{hanhart2000} includes more amplitudes than our analysis, but
the calculations predict little asymmetry for $A_y(\theta)$. 
The differences for $A_{y}$ and the increasing divergence with energy 
are also  seen in Fig.~15. 
At this time there are no predictions available for $A_{xz}$ and $A_{z0}$.
%
\vspace{-0.3cm}
\hspace{-2.cm}
\begin{figure}[htb]
\epsfxsize=8.55cm
\centerline{
\epsfbox{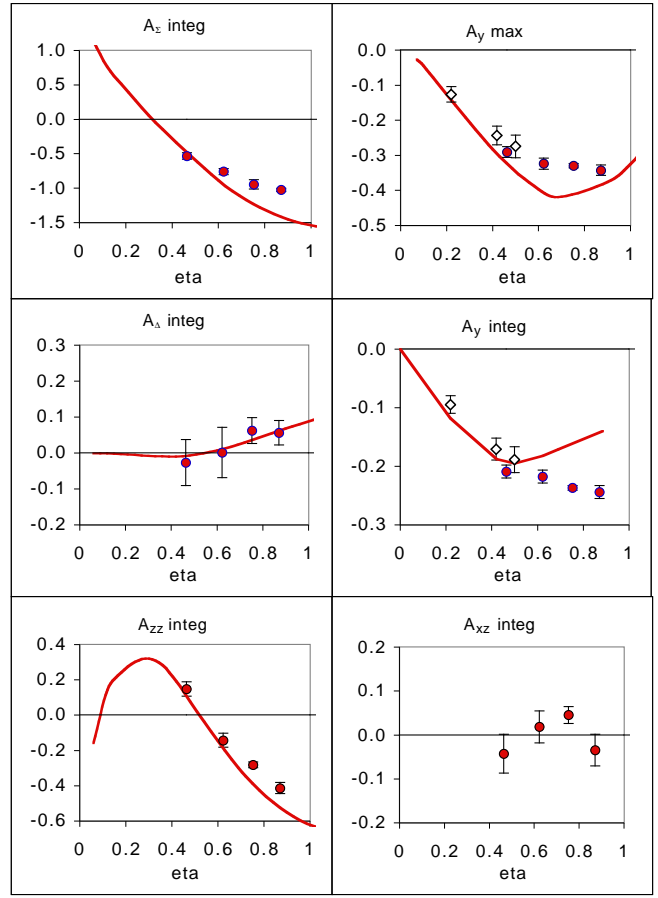}
}
\vspace{0.25cm}
\caption{Energy dependence of the integrated spin correlation coefficients
$\overline{A_{\Sigma}}$, $\overline{A_{\Delta}}$, $\overline{A_{zz}}$  
$\overline{A_{xz}}$ and $\overline{A_{y0}}$, and the peak 
analyzing power $A_{y,max}$ 
for {$\vec{p}\vec{p}\rightarrow pn\pi^+$}.
The diamond shape symbols represent measurements at lower energies and are 
taken from Ref.~[18].  
The solid lines are predictions of the J\"ulich meson exchange model. 
(There is no prediction for $A_{xz}$.)}
\label{fig:Aij(E)15}
\end{figure}
%
\subsection {Deduction of important partial waves}

The number of contributing partial waves grows rapidly with 
energy. If we restrict ourselves to Ss, Sp, Ps, and Pp 
contributions as above, the 19 individual amplitudes listed 
in Table IV are needed for a detailed interpretation of the data.
There are 12 isoscalar and 7 isovector amplitudes. 
The experimental information available includes the three
 cross sections $\sigma_{tot},\ \Delta \sigma_{T}$, 
and $\Delta \sigma_{L}$ for \ppn that are related to these 
amplitudes. 
In addition a recent \ppp study  
provides the three relevant isovector cross sections
$\sigma'_{tot},\ \Delta\sigma'_{T}$, 
and $\Delta\sigma'_{L}$ (Ref.~\cite{meyer01}, Table V). 
As long as isospin is a good quantum number these cross sections 
also give the isovector part of the \ppn reaction if taken at
the same $\eta$. 
\onecolumn 
\narrowtext 
So one has 6 new measurements for 19 variables. 
This necessitates some restriction of the further analysis.
In a previous \ppn study \cite{flammang}, closer to threshold, the partial 
wave space had been restricted to the lowest isoscalar amplitudes $a_{0},
a_{1}, a_{2}$, and to the lowest known isovector amplitudes.
With this simplification and with reliance on the measured 
analyzing powers three amplitudes were deduced for 
$\eta \leq 0.5$. 
Some of these earlier results will be shown below. 
It will become apparent in comparison with our new data 
that the angular momentum space considered in Ref.\cite{flammang} 
is too small for $\eta > 0.3$.
For $0.3 <\eta < 0.9$ it becomes necessary to consider all Ss, Sp, Ps, 
and Pp contributions. In order to 
reduce the number of variables we use similarities in the 
spin algebra coefficients for the 19 amplitudes of interest. 
A suitable combination of the 19 partial cross sections into six groups
allows us to find the Sp and Ps strengths separately 
to deduce the lowest Pp isoscalar partial cross section for the amplitude 
$a_{3}$ directly, and to put a close upper limit on the Ss contributions.
We will identify the isoscalar partial wave cross sections by 
$\sigma(a_{0}),\ \sigma(a_{1}),\ \sigma(a_{2}),\ \ldots$ and 
the isovector partial wave cross sections by
$\sigma(b_{0}),\ \sigma(b_{1}),\ \sigma(b_{2}),\ \ldots$ as in Table V.
Generally, $\sigma(a_{i})=C_{i}|a_{i}|^{2}$, where the $C_{i}$ factor is 
a combination of factors like $\pi$ and Clebsch-Gordan coefficients, 
which can differ from amplitude to amplitude. 
\noindent (Therefore the partial 
cross sections listed in Table V  do not 
provide the magnitude of the corresponding amplitudes without further 
 work.)
The notation $\sigma(a_{1},a_{4\rightarrow 6})$ implies that we could not 
separate the cross sections for $a_{1}$, the Ss component, from the Pp 
components $a_{4}$ to $a_{6}$. Hence $\sigma(a_{1},a_{4\rightarrow 6}) 
\equiv  \sigma(a_{1})+\sigma(a_{4})+\sigma(a_{5})+\sigma(a_{6})$.
The partial cross section groups that could be isolated are given in 
Eq.~\ref{amplitudes}:
%
\widetext 
\begin{mathletters}
\label{amplitudes}¶
\begin{eqnarray}
{\rm Sp\ isoscalar\ terms:}\ \ \ \ \ \ \  \sigma(a_{0},a_{2}) & = & {\frac{1}{8}} \left ({\Delta \sigma_{L}}+2{\Delta \sigma_{T}}+2{\sigma_{tot} }
-{\Delta\sigma_{L}'}-2{\Delta\sigma_{T}'}-2{\sigma'_{tot}} \right) \\[1.2ex]
{\rm Ps\ isovector\ terms:}\ \ \ \ \ \ \ \  \sigma(b_{1},b_{2}) & = & 
 {\frac{1}{8}} \left ({\Delta\sigma_{L}'} +2{\Delta\sigma_{T}'} +2{\sigma'_{tot}} \right)\\[1.2ex]
{\rm Ss+Pp\ isoscalar\ terms:}\ \ \ \ \sigma(a_{1},a_{4\rightarrow 6}) & = &  {\frac{1}{4}} \left (-{\Delta 
\sigma_{L}} +2\sigma_{tot} 
+ {\Delta\sigma_{L}'} -2{\sigma'_{tot}}\right) \\[1.2ex]
{\rm Ss+Pp\ isovector\ terms:}\ \ \ \ \ \ \ \  \sigma(b_{0},b_{3}) & = & {\frac{1}{8}} \left ( 
{\Delta\sigma_{L}'} -2{\Delta\sigma_{T}'} +2{\sigma'_{tot}}\right) \\[1.2ex]
{\rm Pp\ isoscalar\ terms:}\ \ \ \ \ \ \ \ \ \ \ \  \sigma(a_{3}) & = &{\frac{1}{8}} \left ( {\Delta \sigma_{L}}-2{\Delta \sigma_{T}}+2\sigma_{tot} 
- {\Delta\sigma_{L}'} +2{\Delta\sigma_{T}'} -2{\sigma_{tot}'}\right) \\[1.2ex]
{\rm Pp\ isovector\ terms:}\ \ \ \ \ \ \  \sigma(b_{4\rightarrow 11}) & = & {\frac{1}{4}} \left (
-{\Delta\sigma_{L}'} +2{\sigma'_{tot}} \right)\\
%
\nonumber
\end{eqnarray}
\end{mathletters}
\narrowtext
Of these six equations, which hold for \ppn, three also hold for \ppp.
We note that Eq.~\ref{amplitudes}b has been presented before. 
It is identical to Eq.~13 in 
\cite{meyer01}. The six equations now permit a calculation of partial 
wave cross sections to the specified groups of final states
\twocolumn 
\noindent
from the measured spin-dependent cross sections.
The sum of these partial cross 
sections equals the total  $\pi^{+}$ production cross section.
Since the partial cross sections add incoherently the effect of higher 
lying weak amplitudes is minimized. This is an advantage over 
relying on analyzing powers, which are sensitive to even small admixtures.
The amplitudes included in each group are indicated on the left 
side of Eqs.~\ref{amplitudes}. 

In many experiments, including the present one, it is much easier to 
measure accurate cross section ratios than absolute cross sections. 
So our experimental \ppn quantities are given as a fraction of the 
total $\pi^{+}$ production cross section $\sigma_{tot}$.
The Eqs.~\ref{amplitudes} are easily re-written in terms of partial wave 
strengths by dividing both sides by ${\sigma_{tot}}$.
For the $p+\pi^{+}$ branch the values 
${\Delta\sigma_T}/{\sigma_{tot}}$ and ${\Delta\sigma_L}/{\sigma_{tot}}$ 
were calculated from eqs.~4.6 and 
$A_{\Sigma}(\theta)$, $A_{\Delta}(\theta)$ and $\sigma_{0}(\theta)$. 
In figures and tables we will generally use the ratios of partial wave 
cross sections to total cross sections. We refer to them as partial 
wave strengths.
%
\vspace{-0.4cm}
\hspace{-0.cm}
\begin{figure}[htb]
\epsfxsize=7.cm
\centerline{
\epsfbox{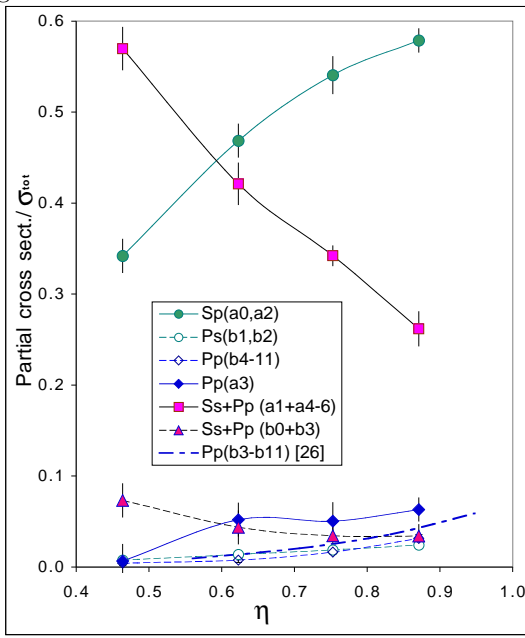}
}
\vspace{0.20cm}
\caption{Partial wave strengths for six groups of amplitudes as function of 
$\eta$. The isoscalar cross sections are connected by solid lines, 
the isovector ones by dashed lines. The contributing amplitudes, including 
the (small) Pp contributions not resolved from the dominant Ss 
cross sections are indicated in the legend. The dash-dotted line 
represents the full Pp isovector strength contribution in \ppn, as
derived from the results of Ref. [26].
}
\label{fig:AmplitudesSig}
\end{figure}
To use them in this study the total \ppp 
and \ppn cross sections were taken from the literature and interpolated
for the present $\eta$ values. We obtained the \ppp information needed 
from Ref.~\cite{meyer01} and the \ppn total cross sections 
from \cite{flammang} and from Fig.~2 in Ref.~\cite{daehnick95}. 
The accuracy of the total cross section ratios so obtained is not very
high, but it will suffice here because the isoscalar terms of interest
are an order of magnitude larger than the isovector terms.
The partial cross section strengths derived with Eqs. 
\ref{amplitudes} are displayed in Fig.~16 and listed in in Table V.

The primed  cross sections are the (pure isovector)
cross sections measured for \ppp, which are also more accurately 
given as fractional strengths. To work in terms of \ppn partial wave 
strengths the \ppp strengths of Ref.~\cite{meyer01} 
have to be multiplied by the ratio of the \ppp and \ppn unpolarized 
cross sections, taken at the same relevant $\eta$ values. 

Fig.~\ref{fig:AmplitudesSig} shows the change of partial wave 
strength with energy for Sp, Ps, and other groups.
It is immediately apparent that for the energy region studied the 
leading isoscalar partial cross sections are an order 
of magnitude larger than the isovector ones.
It helps our discussion that lowest-lying 
Pp isoscalar partial wave strength Pp($a_{3}$)
could be resolved. It is much smaller than the Ss and Sp 
strengths. So is the sum of all isovector cross sections for
$\rm b_{4}\ to\ b_{11}$.
The Pp strengths attributable to $b_{3}$ can be assessed by comparing 
Pp($b_{4 \rightarrow 11}$) from this work with the heavy dash-dotted 
curve derived from \cite{meyer01} for the full Pp isovector strength 
Pp($b_{3 \rightarrow 11}$).
\vspace{0.10cm}
\vspace{-0.5cm}
\hspace{-0.cm}
\begin{figure}[htb]
\epsfxsize=7.cm
\centerline{
\epsfbox{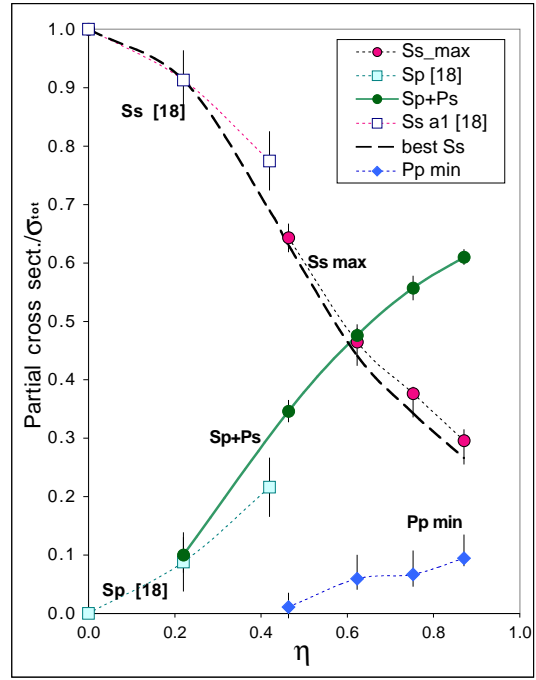}
}
\vspace{0.20cm}
\caption{Sums of isoscalar and isovector partial wave strengths as 
function of $\eta$. The Sp+Ps sum is measured directly. The points 
labeled  ``Ss max'' represent a close upper limit to the sum of the 
Ss partial cross sections. Any correction for 
the unresolved Pp amplitudes ($a_{4}, a_{5}, a_{6}$, and $b_{3}$) would  
lower the Ss curve (as indicated by the estimated errors). The 
admixtures can be expected to be smaller than $Pp(a_{3}$). The lower 
points show the documented Pp strengths only. The data at $\eta=0.22$ and 
0.42 are from Ref.~18.
}
\label{fig:pnpartials}
\end{figure}
Therefore, it seems reasonable to assume that the Pp contributions from 
$a_{4},\ a_{5}, a_{6}$ and $b_{3}$, which could not be disentangled 
from the the Ss amplitudes, are also much smaller than the Ss terms. 
On this basis we estimate that they make up no more than 5 to 10\% 
of the ``Pp entangled'' $\rm Ss_{max}$ cross section curve.
For $\eta=0.9$ the Sp($a_{0},a_{2})$ cross section has become 
dominant. As seen in 
Fig.~16, it is very much larger than the Ps isovector contribution.
It would be of interest to resolve the isoscalar component
$a_{0}$, because it can be used to constrain the strength of 3-body 
forces \cite{hanhart2000b}.
However, in this analysis $a_{0}$ and the much larger amplitude $a_{2}$ 
always appear together. 
The Ss fraction, including
the unresolved Pp contributions, has fallen to less than 0.3. This is 
consistent with the work at 420 MeV \cite{falk}. 

In Fig.~\ref{fig:pnpartials} the data points give the summed Sp+Ps 
strengths, the upper limit for the summed Ss strengths, and a lower 
limit for the Pp strength. The heavy dashed curve shows the likely 
energy dependence of the actual Ss strength. 
The divergence of the old and new \ppn interpretation
near $\eta\approx 0.45$ serves as a reminder that a partial wave 
analysis is only model independent if it fully encompasses all 
contributing amplitudes. This apparently was 
no longer true for the 320 MeV data ($\eta=0.42$) of Ref.~18. 

In this respect our present difficulty to perfectly reproduce  
$A_{y}$ at 375 and 400 MeV  in the Ss, Sp, Ps and Pp frame work 
(see Table II) should be taken as a warning. At these energies 
some higher partial waves may contribute enough 
so that they must be considered, at least for the analyzing powers.

\section{Conclusions}

We have measured the spin correlation coefficients
$A_{\Sigma}$=$A_{xx}$+$A_{yy}$, 
$A_{\Delta}$=$A_{xx}$-$A_{yy}$, $A_{zz}, A_{xz}$, and $ A_{y}$, as well as 
 angular distributions for $\sigma(\theta_{\pi})$ and the polarization 
 observables $A_{ij}(\theta_{\pi})$ at energies from 325 to 400 MeV. 
At the lowest energies the results are in 
agreement with prediction of the J\"ulich meson exchange model.
The agreement deteriorates considerably at energies where Ss 
transitions no longer dominate. At 375 and 400 MeV some physics 
aspects in {$\vec{p}\vec{p}\rightarrow pn\pi^+$} apparently
are missed by the model. This suspicion is 
supported by the even poorer agreement of the model with the
{$\vec{p}\vec{p}\rightarrow pp\pi^0$} data \cite{meyer01}.

The \ppp \ and \ppn \ reactions are found to differ greatly 
in the relative importance of Sp, Ps, and Pp transitions. Sp strongly 
feeds the delta resonance in \ppn, but this transition is forbidden 
for {$\vec{p}\vec{p}\rightarrow pp\pi^0$}. By contrast Ps contributions
in \ppn are no larger than Pp contributions, as seen in Fig.~16.
In {$\vec{p}\vec{p}\rightarrow pn\pi^+$} the Ss and Sp isoscalar
terms are most important while the 
Pp transitions just begin to contribute. 
For {$\vec{p}\vec{p}\rightarrow pp\pi^0$} Pp 
becomes dominant at $\eta=0.7$. 
  
The partial wave analysis was able to reproduce almost all
 polarization observables within experimental errors. 
This supports the postulated adequacy of considering only 
Ss, Sp, Ps, and Pp transitions in the near threshold 
region. However, this angular momentum space 
may not be adequate to explain details of analyzing powers, because 
they can be affected by small admixtures of higher-lying transitions.
Even in this limited space the number of individual partial 
waves for {$\vec{p}\vec{p}\rightarrow pn\pi^+$}
at 400 MeV is too large to deduce all individual amplitudes. 
Some interesting sum rules for groupings of amplitudes were found 
(Eqs.~\ref{amplitudes}), and the corresponding partial 
cross sections could be extracted. They show, e.g., 
that for $\eta < 1$ Pp and Ps terms play a considerably 
smaller role in \ppn than in \ppp.

Further progress may come from improved theoretical models 
that can accurately predict the new data at hand. It is interesting 
to note again that the J\"ulich model does well at 325 MeV where 
Ss dominates, but it increasingly fails for \ppn  
(as well as for \ppp) as higher angular momenta become important.

\section{Acknowledgements}

We acknowledge the assistance of Drs.~M.~Dzemidzic, F.~Sperisen 
and D.~Tedeschi in the early stages of the experiment. Throughout the runs
we have benefitted from the helpful advice of Dr. W. Haeberli 
and the technical assistance of J. Doskow. We  
thank the IUCF accelerator operations group for their
dedicated efforts. We are grateful to the authors of
ref. \cite{hanhart2000} for making available to us calculations
for $\vec{p}\vec{p}\rightarrow pn\pi^+$ obtained with their model.

This work was supported by the US National Science Foundation under
Grants PHY95-14566, PHY96-02872, PHY-97-22556, PHY-99-01529, and by 
the department of Energy under Grant DOE-FG02-88ER40438.

\onecolumn 

\begin{table}
\caption{Beam energies, integrated luminosities for the p + $\pi^{+}$ 
measurements, and the products of beam and target polarization 
for runs a and b. (No p+n data were taken in run a. The p+n measurements began 
in the middle of run b and have correspondingly lower integrated luminosities.)}
\begin{tabular}{c |l c  |l ccc}
    
     	       &\ \ \ \ Run a & & &\ \ \ \ \ \ \ \ \ Run b && \ \ \   \\
\tableline   
Energy &$\int L\,dt$&$P_y Q$ & $\int L\,dt$ &$P_x Q$ &$P_y Q$ &$P_z Q$   \\
 (MeV) &($nb^{-1}$) &        & ($nb^{-1}$)  &        &        &         \\
\tableline  
325.6 &2.163 &0.456 $\pm$0.003 &3.0  &0.059 $\pm$0.002 &0.333 $\pm$0.002 &0.296 $\pm$0.003 \\
350.5 &0.901 &0.342 $\pm$0.004 &1.3  &0.053 $\pm$0.003 &0.316 $\pm$0.003 &0.267 $\pm$0.005 \\
375.0 &3.024 &0.514 $\pm$0.004 &4.1  &0.041 $\pm$0.002 &0.333 $\pm$0.002 &0.266 $\pm$0.004 \\
400.0 &0.831 &0.526 $\pm$0.006 &1.1  &0.039 $\pm$0.004 &0.289 $\pm$0.004 &0.203 $\pm$0.008
\end{tabular}
\end{table}

\begin{table}
\caption{Coefficients for the fits with Eqs.~\ref{fit_eq}
that reproduce the measured angular distributions of the polarization observables.
The associated Legendre polynomials used for the fits are determined by selection 
rules for Ss, Sp, Ps, and Pp transitions. The unpolarized angular 
distribution $\sigma_{0}(a_{00},b_{00})$ is given in arbitrary 
units by setting $a_{00}=1$.
The errors listed refer to the individual fitting coefficients. The $\chi^{2}$ 
numbers give the overall quality of the fit to the data per degree of freedom.
The fits are shown in Figs. 11, 13 and 14.}
\begin{tabular}{c |l c c |l c c |l c c |l c c}
 
             &      & 325 MeV &  &  & 350 MeV &  &   &    &375 MeV &    &   & 400 MeV  \\
   Name      &param.&param.&$\chi^{2}$&param.&param.&$\chi^{2}$&param.&param.&$\chi^{2}$&param.&param.&$\chi^{2}$\\    
             &value &error &data  & value &error &data    &value &error&data  & value &error&data\\
 \tableline   
$a_{00}     $ &1     & -   & -  & 1     &  - &  -  & 1    &  -  &  - & 1    &  -  & -\\ 
$b_{00}     $ &0.168 &0.035& -  & 0.190 &0.040&  - & 0.199&0.030& -  &0.196 &0.045& \\
 \tableline 
$a_{\Sigma}$ &-0.560&0.052&0.5 &-0.810 &0.055&0.6 &-0.994&0.015&2.8 &-1.070&0.024&1.4 \\ 
$b_{\Sigma}$ &-0.303&0.063&    &-0.478 &0.067&    &-0.510&0.018&    &-0.439&0.029& \\ 
 \tableline 
$a_{\Delta}$ &-0.037&0.091&1.4 &-0.001 &0.097&0.2 &0.084 &0.028&1.3 & 0.075&0.045&1.4\\ 
$b_{\Delta}$ & -    &  -  &  - &   -   & -   &  - & -    &   - &  - &-     & -   &  \\ 
 \tableline 
$a_{zz}    $ & 0.120&0.042&0.1 &-0.177 &0.047&0.8 &-0.310&0.018&1.0 &-0.431&0.037&3.1 \\ 
$b_{zz}    $ &-0.188&0.054&    &-0.257 &0.057&    &-0.233&0.021&    &-0.199&0.043&  \\ 
 \tableline 
$a_{y0}    $ &-0.247&0.015&0.5 &-0.255 &0.013&0.9 &-0.276&0.005&2.4 &-0.285&0.008&3.9 \\ 
$c_{y0}    $ & 0.007&0.013&    & 0.050 &0.010&    &0.044 &0.005&    &-0.032&0.007&  \\ 
 \tableline 
$a_{xz}    $ &-0.051&0.042&0.7 & 0.021 &0.042&1.2 &0.053 &0.021&1.1 &-0.041&0.041&1.2 \\ 
$c_{xz}    $ & 0.106&0.038&    & 0.047 &0.036&    &-0.040&0.020&    &-0.104&0.036& 

\end{tabular}
\end{table}

\begin{table}
\caption{Beam energy, the $\eta$ parameter, and the 
deduced integrated spin correlation coefficients. The table gives the 
weighted average of all runs as shown in Fig.~15.}
\begin{tabular}{ccccccc}
	      T(MeV)&$\eta$& $\overline{A_{\Sigma}}$ & 
	      $\overline{A_{\Delta}}$ &$\overline{A_{zz}}$ & $\overline{A_{y0}}$ & $\overline{A_{xz}}$  \\
\tableline
325.6&0.464&-0.533 $\pm$0.046 &-0.027 $\pm$0.064& 0.148 $\pm$0.041 &-0.209 $\pm$0.011 &-0.043 $\pm$0.044\\
350.5&0.623&-0.761 $\pm$0.046 & 0.001 $\pm$0.070&-0.143 $\pm$0.040 &-0.218 $\pm$0.011 & 0.018 $\pm$ 0.036\\
375.0&0.753&-0.945 $\pm$0.068 & 0.062 $\pm$0.036&-0.283 $\pm$0.016 &-0.237 $\pm$0.004 & 0.045 $\pm$0.019\\
400.0&0.871&-1.026 $\pm$0.023 & 0.056 $\pm$0.034&-0.414 $\pm$0.032 &-0.244 $\pm$0.011 &-0.035 $\pm$0.036
\end{tabular}
\end{table}
\narrowtext
\begin{table}
\caption{Angular momentum quantum numbers for the  partial
waves of the reaction \ppn. 
}
\label{tab4}
\begin{tabular}{c c c c }
Type & & Label & $^{2s_i+1}l_J \rightarrow  \kern 1pt ^{2s_f+1}l_{p}j,l_{\pi} $ \\
\tableline
{\bf Ss}&isoscalar & $a_{1}$ &$^3P_1\rightarrow ^3\!S_1, s$ \\
{\bf Ss}&isovector & $b_{0}$ &$^3P_0\rightarrow ^1\!S_0, s$ \\
\hline
\bf Sp  &isoscalar & $a_{0}$ &$^1S_0\rightarrow ^3\!S_1,p$ \\
        &          & $a_{2}$ &$^1D_2\rightarrow ^3\!S_1,p$ \\ 
\hline	
\bf Ps  &isovector & $b_{1}$ &$^1S_0\rightarrow ^3\!P_0,s$ \\
        &          & $b_{2}$ &$^1D_2\rightarrow ^3\!P_2,s$ \\
\hline
\bf Pp  &isoscalar & $a_{3}$ &$^3P_0\rightarrow ^1\!P_1,p$ \\
        &          & $a_{4}$ &$^3P_1\rightarrow ^1\!P_1,p$ \\ 
        &          & $a_{5}$ &$^3P_2\rightarrow ^1\!P_1,p$ \\ 
        &          & $a_{6}$ &$^3F_2\rightarrow ^1\!P_1,p$ \\ 
\hline
\bf Pp	&isovector & $b_{3}$ &$^3P_0\rightarrow ^3\!P_1,p$ \\
        &          & $b_{4}$ &$^3P_2\rightarrow ^3\!P_1,p$ \\
        &          & $b_{5}$ &$^3P_2\rightarrow ^3\!P_2,p$ \\
        &          & $b_{6}$ &$^3F_2\rightarrow ^3\!P_1,p$ \\
        &          & $b_{7}$ &$^3F_2\rightarrow ^3\!P_2,p$ \\
        &          & $b_{8}$ &$^3P_1\rightarrow ^3\!P_0,p$ \\
        &          & $b_{9}$ &$^3P_1\rightarrow ^3\!P_1,p$ \\
	&          & $b_{10}$&$^3P_1\rightarrow ^3\!P_2,p$ \\
	&          & $b_{11}$&$^3F_3\rightarrow ^3\!P_2,p$ \\
\end{tabular}
\end{table}

\mediumtext    
\begin{table}
\caption{ Listing of the \ppn partial wave strengths for the  groups of 
isoscalar and isovector amplitudes indicated. The 300 MeV results listed were 
taken from Ref.~[18]. The 300 MeV strengths not listed are assumed to 
be negligible.  
}
\label{tab5}
\begin{tabular}{l c|c  c |c  c |c  c }	
 Isoscalars   \\	
E(MeV)	& \ \ \ \  $\eta$\ \ \ \ \ & Sp(a0,a2) & \ \ \ \  error\ \ \ \ 
&Ss+Pp (a1,a4-6) &\ \ \ error\ \ \ \  &  Pp(a3) &\ \ \ error\\
\tableline 
300	& 0.220     & 0.088	& 0.030     &       0.740	&	0.050	&	-	& -    \\
325.6	& 0.464     & 0.342	& 0.018     &	0.570	&	0.024	&	0.006	&	0.018  \\
350.5	& 0.623     & 0.469	& 0.018     &	0.421	&	0.023	&	0.052	&	0.018  \\
375	& 0.753     & 0.541	& 0.020     &	0.342	&	0.011	&	0.050	&	0.020  \\
400	& 0.871     & 0.579	& 0.013     &	0.262	&	0.019	&	0.063	&	0.013  \\
\hline		
 Isosvectors 	  \\ 
E(MeV)	& $\eta$ & Ps(b1,b2) &	error	& Ss+Pp (b0,b3)	&	error	&  Pp(b4-11)&	error  \\
\tableline 
300	& 0.220     &  -	& -         &	0.173	&	0.022	&	 -	&	 -    \\
325.6	& 0.464     & 0.007	& 0.002     &	0.073	&	0.007	&	0.004	&	0.003 \\
350.5	& 0.623     & 0.014	& 0.003     &	0.044	&	0.005	&	0.007	&	0.002 \\
375	& 0.753     & 0.019	& 0.002     &	0.034	&	0.003	&	0.016	&	0.002 \\ 
400	& 0.871     & 0.024	& 0.003     &	0.034	&	0.004	&	0.031	&	0.004 \\
\end{tabular}
\end{table}
\end{document}